\documentclass[proof]{pasj00}
\usepackage{rotating}
\begin{document}
\SetRunningHead{M. Kimura et al.}{Suzaku Observations across the Cygnus Loop from Northeastern to the Southwestern Rim }
\title{Suzaku Observations across the Cygnus Loop from the Northeastern to the Southwestern Rim}
\author{Masashi \textsc{Kimura}$^1$, Hiroshi \textsc{Tsunemi}$^1$, Satoru \textsc{Katsuda}$^{1,2}$, Hiroyuki
\textsc{Uchida}$^1$} %
\affil{$^1$Department of Earth and Space Science, Graduate School of
  Science, Osaka University,\\
 1-1 Machikaneyama, Toyonaka, Osaka
  560-0043, Japan}
  \affil{$^2$NASA Goddard Space Flight Center, Greenbelt, MD 20771,
U.S.A.}
  \email{mkimura@ess.sci.osaka-u.ac.jp}

\Received{2008/08/08}
\Accepted{2008/10/21}
\date{}
\makeatletter
\@addtoreset{equation}{section}
\makeatother

\newcommand{\kTe}{$kT_{\rm e}$} 

\maketitle
\begin{abstract}
We have observed the Cygnus Loop from the northeast (NE) rim to the southwest (SW) rim using Suzaku in 10 pointings that is just north  of previous XMM-Newton observations.
The observation data obtained were divided into 45 rectangular regions where the width were configured so that each region holds 8000$-$12000 photons.
The spectrum acquired from each region was fitted either with one-\kTe-component non-equilibrium ionization (NEI) model or with two-\kTe-component NEI model.
The two-\kTe-component model yields significantly better fit in almost all the non-rim regions. Judging from abundances and flux, the high-\kTe-component (0.4$-$0.8keV) must be the ejecta origin, while the low-\kTe-component ($\sim$0.3keV) comes from the swept-up matter.
We found that swept-up matter shell is very thin in just southwest of center of the Loop. Together with previous observations, we estimate the diameter of this thin shell region to be $
1^{\circ}$. 
 We also found that the ejecta distributions were asymmteric to the geometric center: the ejecta of O, Ne and Mg were distributed more in the NE, while the ejecta of Si and Fe were distributed more in the SW of the Cygnus Loop.
We calculated the masses for various metals and estimated the origin of the Cygnus Loop as the $12-15$M$_\odot$ core-collapse explosion.
\end{abstract} 
\newpage
\section{Introduction}
The Cygnus Loop is a nearby (540pc: Blair et al. 2005) middle-aged supernova remnant (SNR). Its large apparent size ($2^{\circ}.5\times3^{\circ}.5$: Levenson et al. 1997; Aschenbach $\&$ Leahy 1999) and high surface brightness enable us to study soft X-ray emission from the Cygnus Loop. 
The bright surface emission is mainly from interstellar medium (ISM) swept-up by forward shock wave.  
Miyata et al. (1994) observed the northeast (NE) rim of the Loop with Advanced Satellite for Cosmology and Astrophysics (ASCA) and revealed metal deficiency there. Since previous measurements indicate that the ISM around the Cygnus Loop has metal deficient abundances (Parker et al. 1967), they concluded that the plasma in the NE rim is dominated by the ISM. 
In the contrast of the rim region, ASCA detected Si, S, Fe rich plasma at the center portion of the Cygnus Loop (Miyata et al. 1998), which is thought to be the ejecta.  
The relative abundances of the ejecta support the idea that the Cygnus Loop was the result from a core-collapse supernova, and the progenitor mass is estimated to be 25M$_\odot$.

\begin{figure*}[htbp]
  \begin{center}
    \FigureFile(80mm,70mm){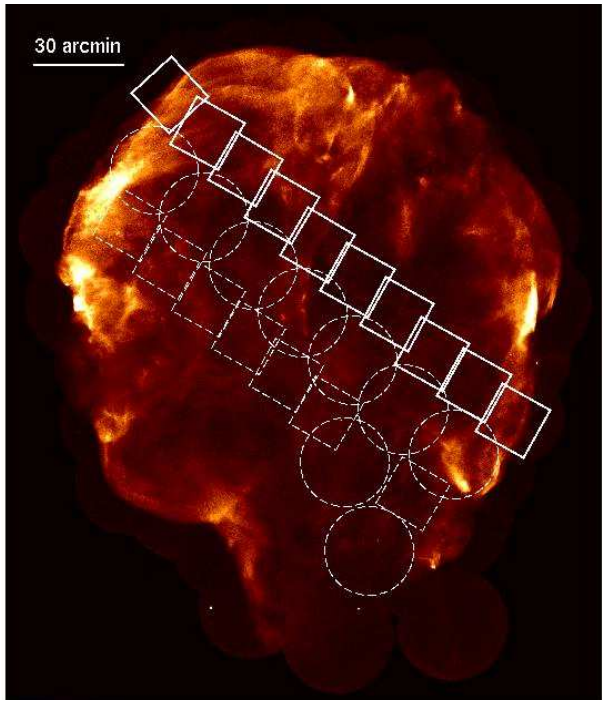}
  \end{center}
  \caption{ROSAT HRI image of the entire Cygnus Loop. The Suzaku FOV (NE3, P1, P2, P3, P4, P5, P6, P7, P9, P10) are shown as white rectangles. Dotted circles and rectangles represent previous XMM-Newton and Suzaku observations.}  
	\label{fig:FOV} 
\end{figure*}

Recent XMM-Newton observations (Tsunemi et al. 2007) across the Cygnus Loop from the NE rim to the southwest (SW) rim revealed ejecta distributed inside of $\sim$0.85 $R_s$ of the Loop, where $R_s$ is the shock radius.
 The relative abundances inferred for the total ejecta are almost consistent with those expected for the core-collapse SN, whose progenitor mass is 13M$_\odot$ (Katsuda \& Tsunemi 2008).
 Suzaku observed just south of XMM-Newton observation path (Katsuda et al. 2008) and revealed asymmetric metal distributions of the Cygnus Loop. They found that Mg is distributed more in the NE, while Si, S, and Fe are distributed more in the SW of the Loop.
In order to extent our knowledge on the ejecta distribution as well as the overlying ISM, we observed the Cygnus Loop from the NE rim to the SW rim using Suzaku observatory. Figure~\ref{fig:FOV} shows our field of view (FOV) which cover just north regions of the XMM-Newton observation path.


\section{Observation and Data Screening}
\begin{figure*}
  \begin{center}
   \includegraphics[height=80mm]{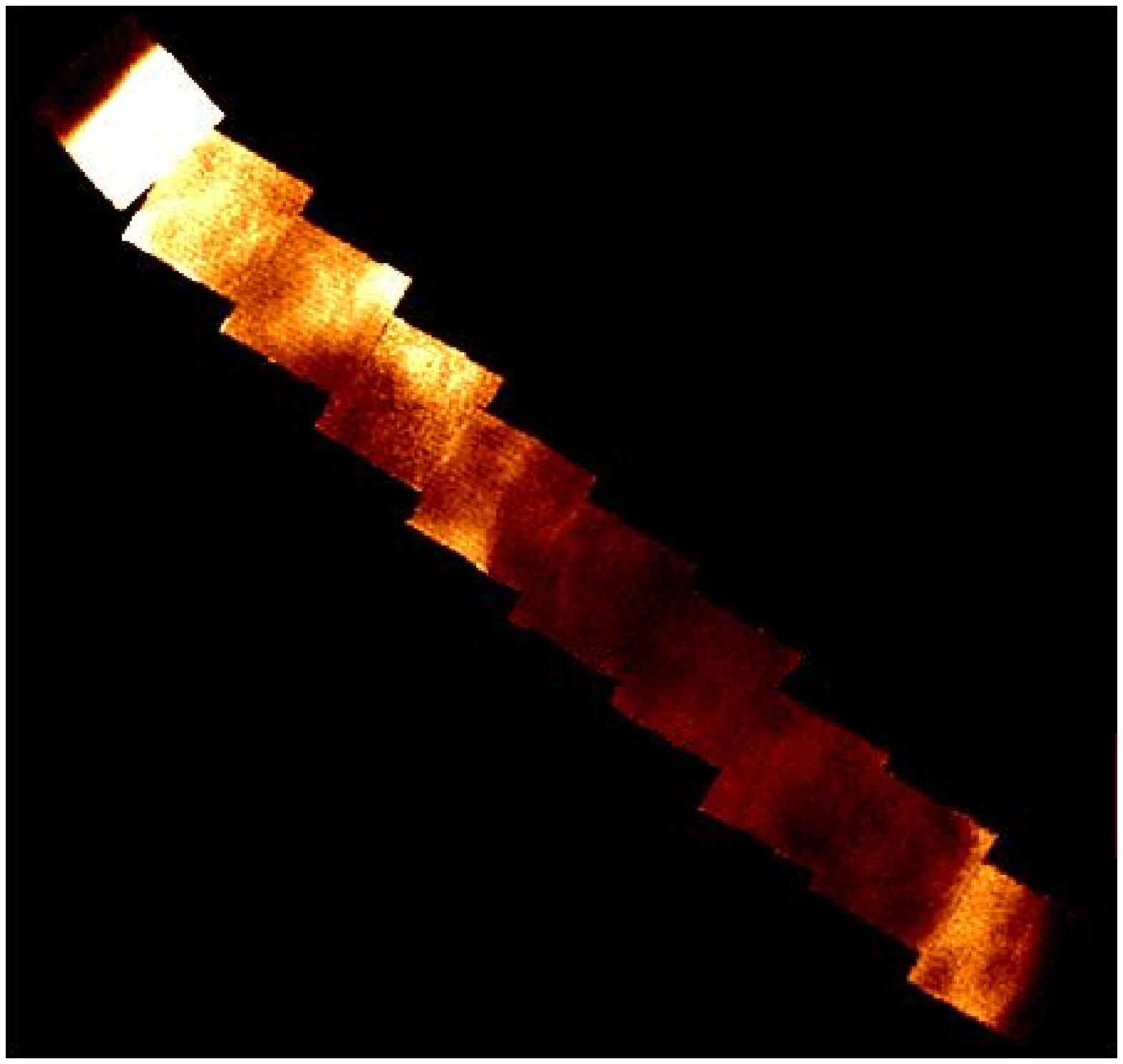}   
    \includegraphics[height=80mm]{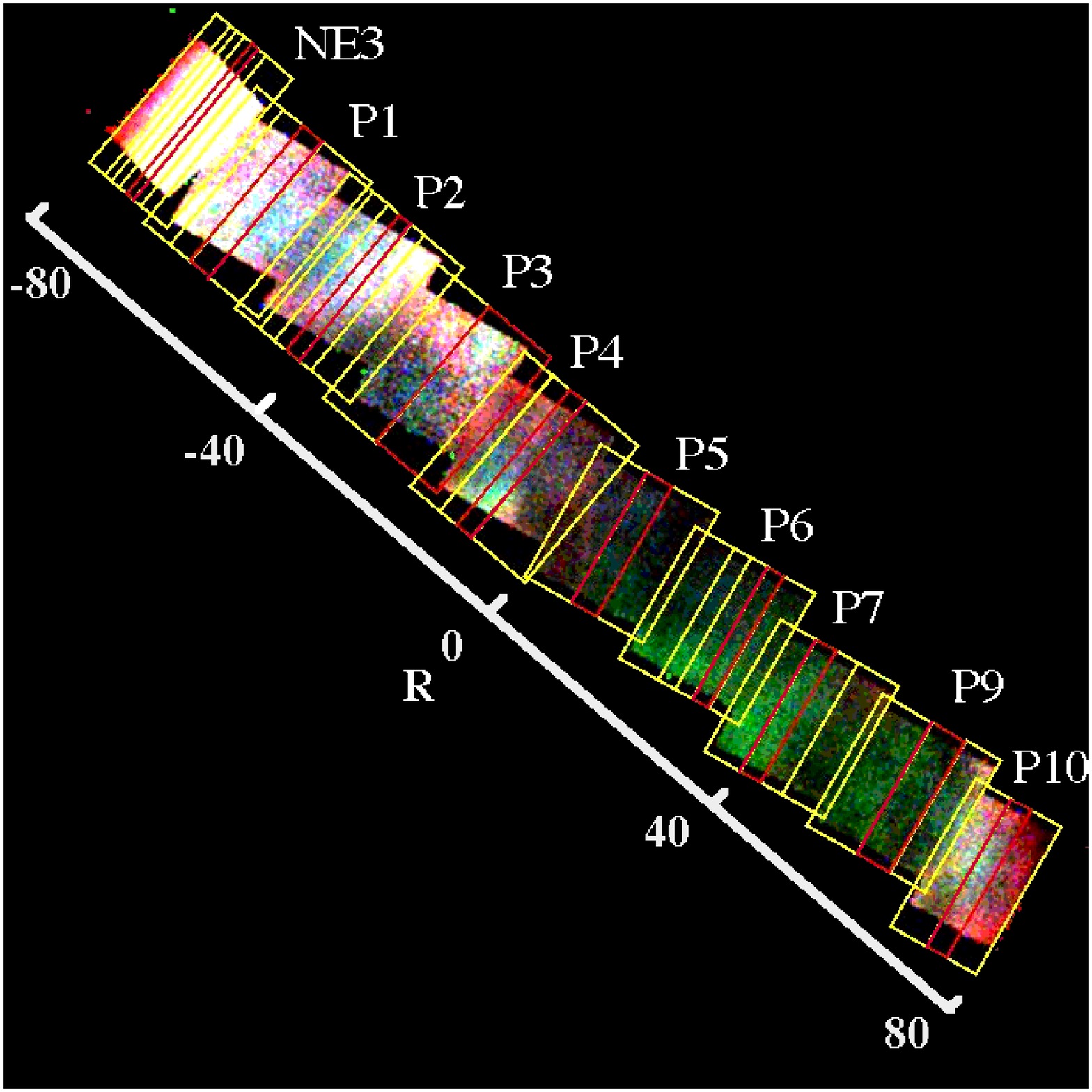}
  \end{center}
  \caption{Left: Surface brightness map of our FOV. Right: Three-color image of ten XIS FOV
 (Red: O {\scshape VII} K$\alpha$, Green: Fe L,
 Blue: Ne {\scshape IX} K$\alpha$).  The data were binned by 8 pixels and smoothed by
 a Gaussian kernel of $\sigma = 25^{\prime\prime}$ The scale R shows the distance from observation center in arcmin. The effects
 of exposure, vignetting, and contamination are corrected for both figure. }  
	\label{fig:xis1_image} 
\end{figure*}

The observations comprised ten points: NE3 , P1$-$P7, P9 and P10.
We employed revision 2.0 of the cleaned event data since there was no sufficient flares in the light curve.
For background subtraction, the spectra acquired from Lockman Hole were used.
We selected the Lockman Hole data whose observation dates were close to those of our Cygnus Loop observations.
Since there was no photons above 3.0 keV after the background subtraction, the energy ranges of 0.3--3.0\,keV and 0.4--3.0\,keV were used for XIS1 (back-illuminated CCD; BI CCD) and XIS0, 3 (front-illuminated CCD; FI CCD), respectively (Koyama et al. 2007).
We summarized Obs IDs, nominal points, observation dates, and effective exposure times after the screening in table~\ref{obs}.
Figure~\ref{fig:xis1_image} right shows a merged XIS1 three-color image.  Red, green, and blue colors correspond to
narrow energy bands of 0.52--0.70\,keV (O {\scshape VII} K$\alpha$),0.70--0.85\,keV (Fe L) and 0.85--0.94\,keV (Ne {\scshape IX} K$\alpha$), respectively.
The scale R shows the distance from an observation center in arcmin.
We can see strong green color around the SW portion (i.e., P6, P7, P9), whereas red and blue colors are enhanced in the NE (i.e., P1$-$P4) regions and P10. 
\section{Spatially resolved spectral analysis}
We divided the entire FOV into 2 parts, namely the NE part and the SW part. The NE part contains pointigs of NE3, P1, P2, P3, P4  and the SW part contains pointings of P5, P6, P7, P9, P10. The two parts were divided into total of 45 rectangular regions where their widths were configured so that each area holds 8000$-$12000 counts of photons in XIS0. The position angle of rectangular regions in the NE part and the SW part are set so that regions are parallel to each rim's shock front. 
The generation of the response matrix file (RMF) and the ancillary response file (ARF),  are done by {\tt xisrmfgen}~(ver. May 2007) and {\tt xissimarfgen}~(ver. March 2008).
In order to reduce the effect of radiation damage of XIS, the data were taken by using the spaced row charge injection (SCI) method (Prigozhin et al. 2008) which recovers the energy resolution of radiation damaged CCD. The SCI method is not fully supported by {\tt CALDB} at the writing phase of this paper, and our data clearly showed better energy resolution than that of the RMF generated by normal use of {\tt xisrmfgen}.
Therefore we searched the appropriate RMF by changing the {\tt date\_obs} parameter of {\tt xisrmfgen} which represents the observation date. 
By changing the {\tt date\_obs} to earlier date, {\tt xisrmfgen} can generate the RMF for earlier observation which has better energy resolution.
We changed {\tt date\_obs} by every 3 months from September 2005 to September 2007, which generate 9 different RMFs in total. Then the generated RMFs were used to fit the 10 sample spectra taken from red regions in figure~\ref{fig:xis1_image}.  We applied model described in next section to test the RMFs in order to find the most appropriate RMF. Since the RMF where we set {\tt date\_obs} = December 2005 showed the least reduced-chi squared value in every spectrum, those RMFs were used to perform further spectral fit.

\subsection{One-\kTe VNEI model}
\begin{figure*}[htbp]
  \begin{center}
    \includegraphics[height=80mm]{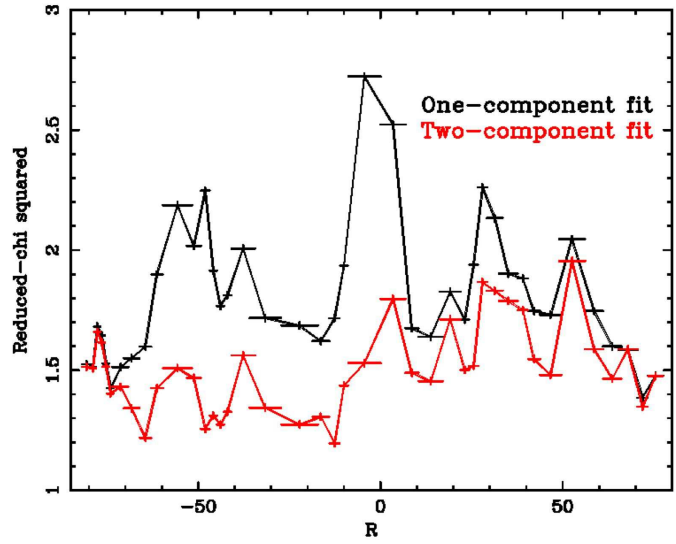}
  \end{center}
  \caption{The comparison of reduced-chi squared of fit between one-component and two-component VNEI model. NE is left.}  
	\label{fig:chi} 
\end{figure*}

The extracted spectra were first fitted by an absorbed non-equailibrium ionization (NEI) model with a single component [the wabs; Morrison \& McCammon 1983 and the VNEI model (NEI ver. 2.0); e.g., Borkowski et al. 2001 in XSPEC v12.4.01].  Free parameters are the hydrogen column density, $N_\mathrm{H}$; electron temperature, \kTe; the ionization time, $\tau$, where $\tau$ is the electron density times the elapsed time after the shock heating; the emission measure EM (EM= $\int n_e n_\mathrm{H} dl$, where $n_e$ and $n_\mathrm{H}$ are the number densities of electrons and hydrogens and $dl$ is the plasma depth); the abundances of C, N, O, Ne, Mg, Si, Fe and Ni. We set abundance of S equal to that of Si and that of Ni equal to that of Fe. Abundances of the other elements were fixed to the solar values (Anders \& Grevesse 1989). 
The black line in figure~\ref{fig:chi} shows the reduced-chi square value as a function of R. The one-component model gave us large reduced-chi square values in some regions; 
therefore, we added an extra component for our spectral fit.
\subsection{Two-\kTe VNEI model}
\begin{figure*}
\begin{center}
   
   \includegraphics[width=70mm]{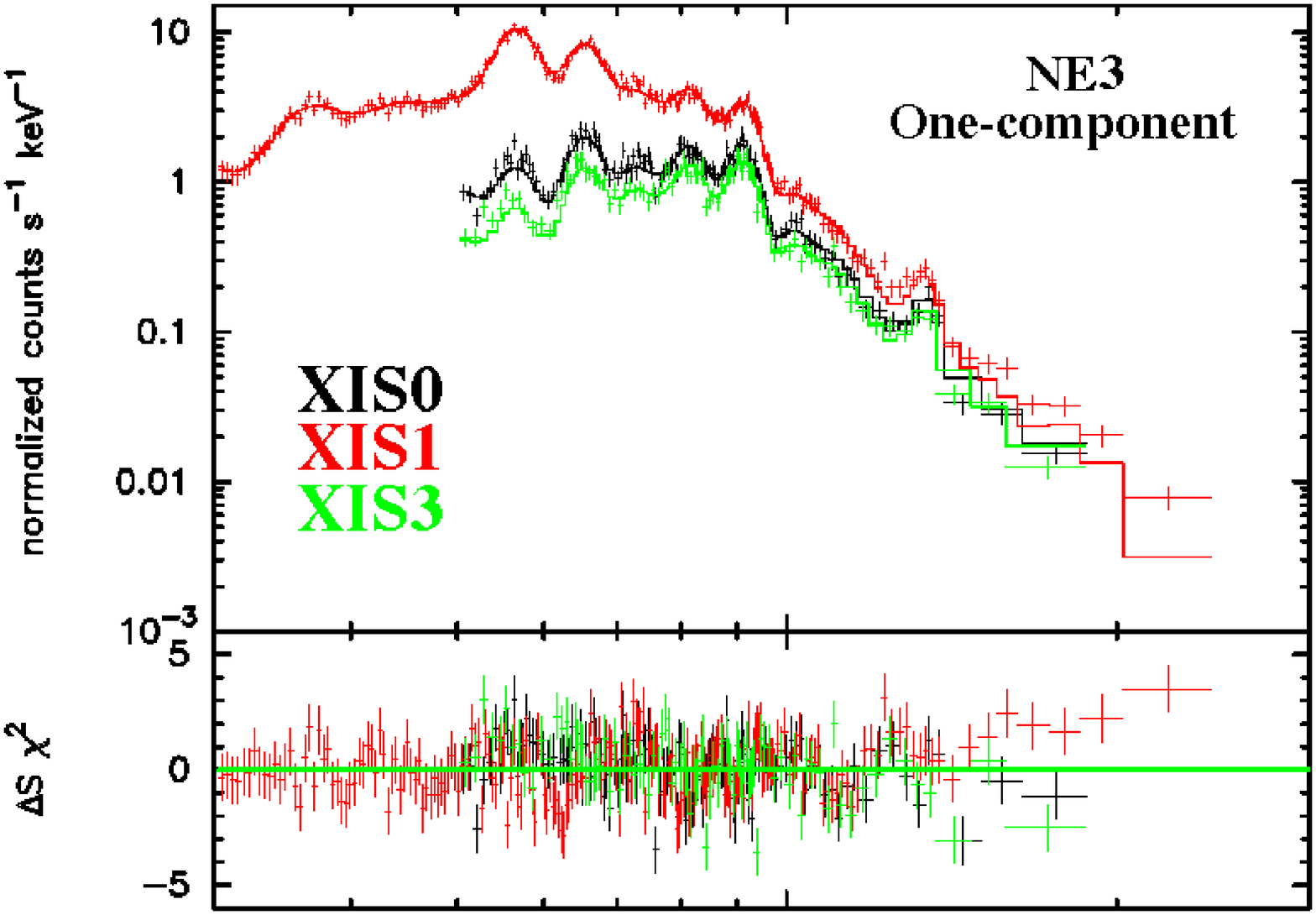}
   \includegraphics[width=70mm]{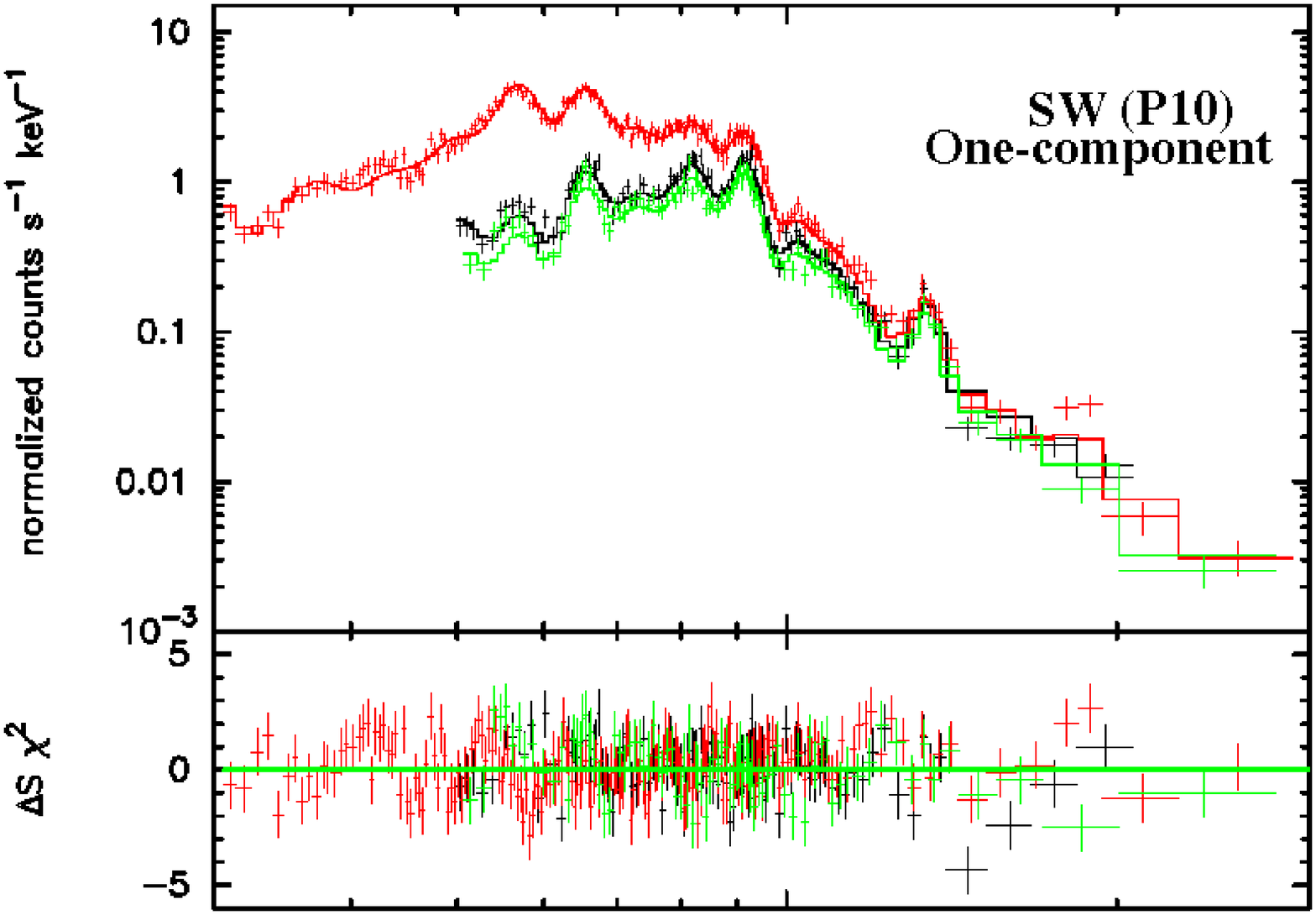}   
   \includegraphics[width=70mm]{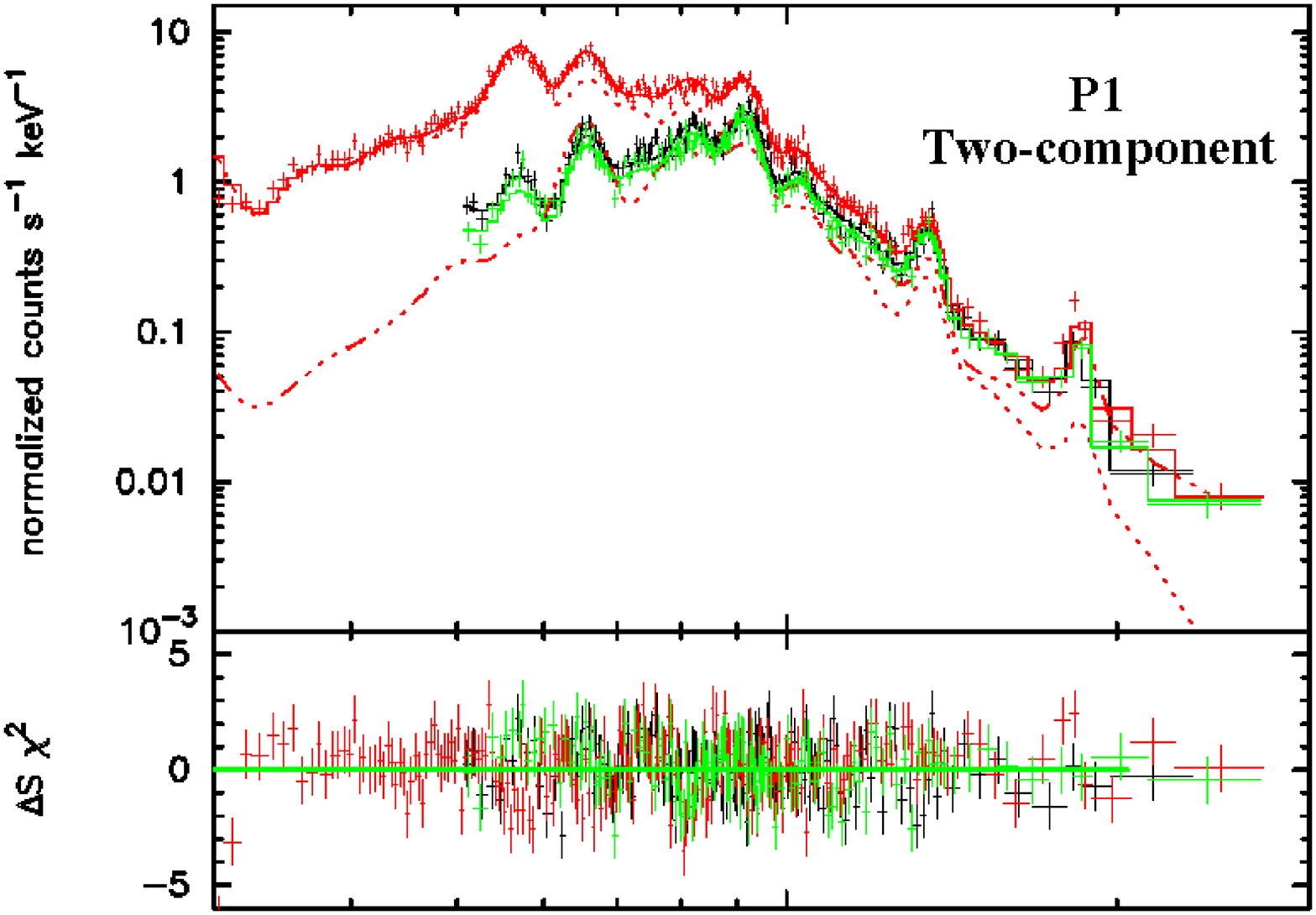}
   \includegraphics[width=70mm]{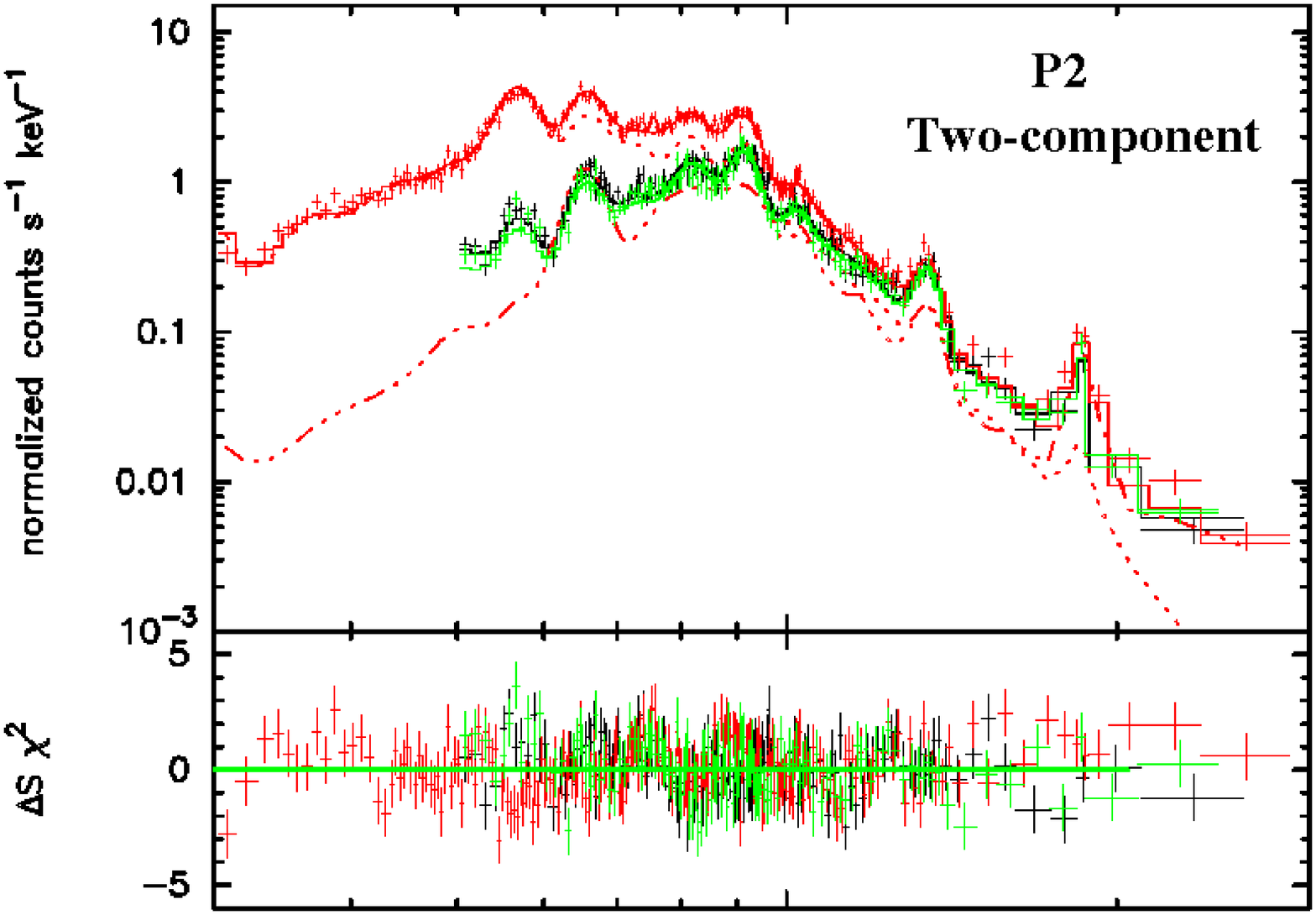}
   \includegraphics[width=70mm]{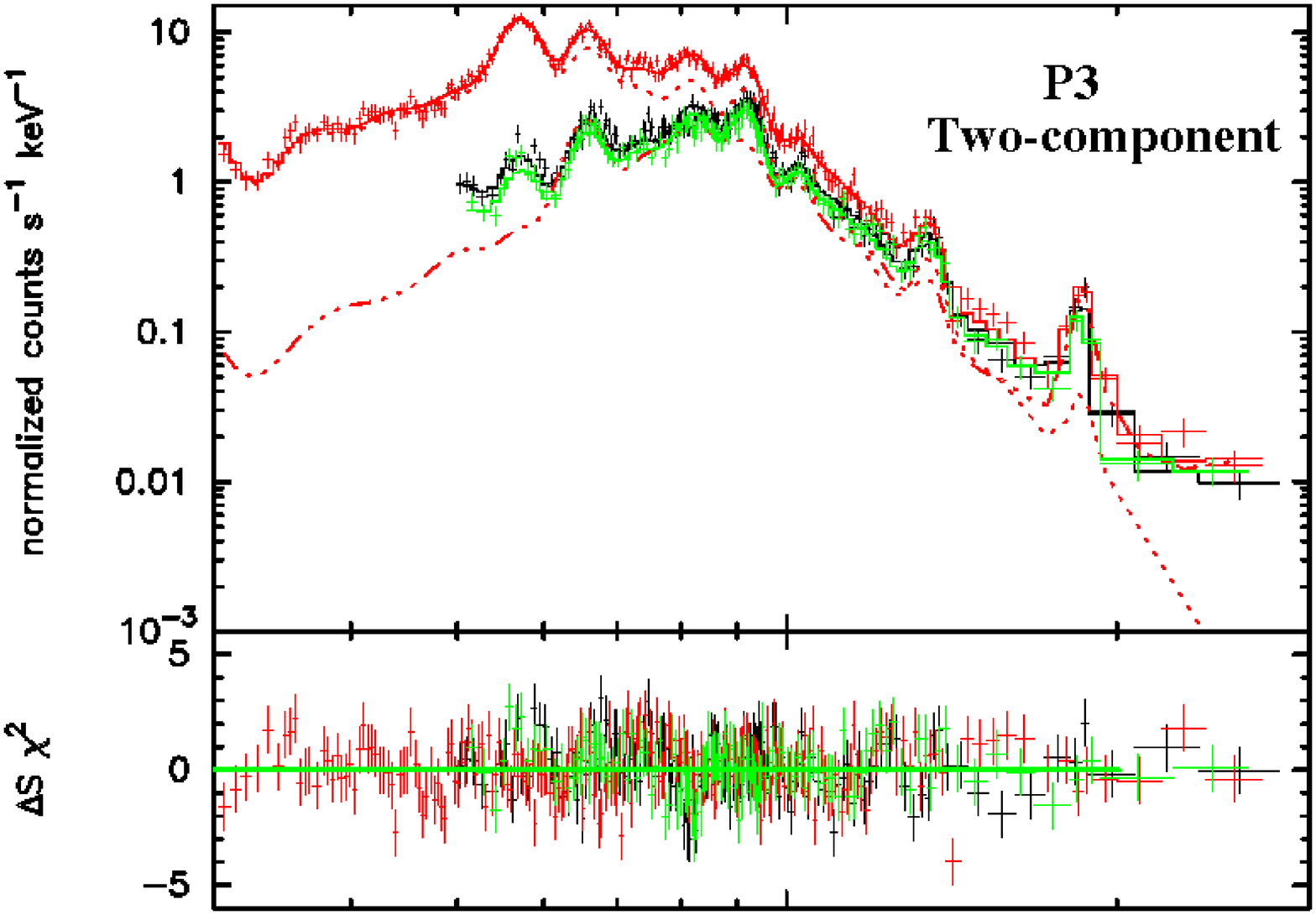}
   \includegraphics[width=70mm]{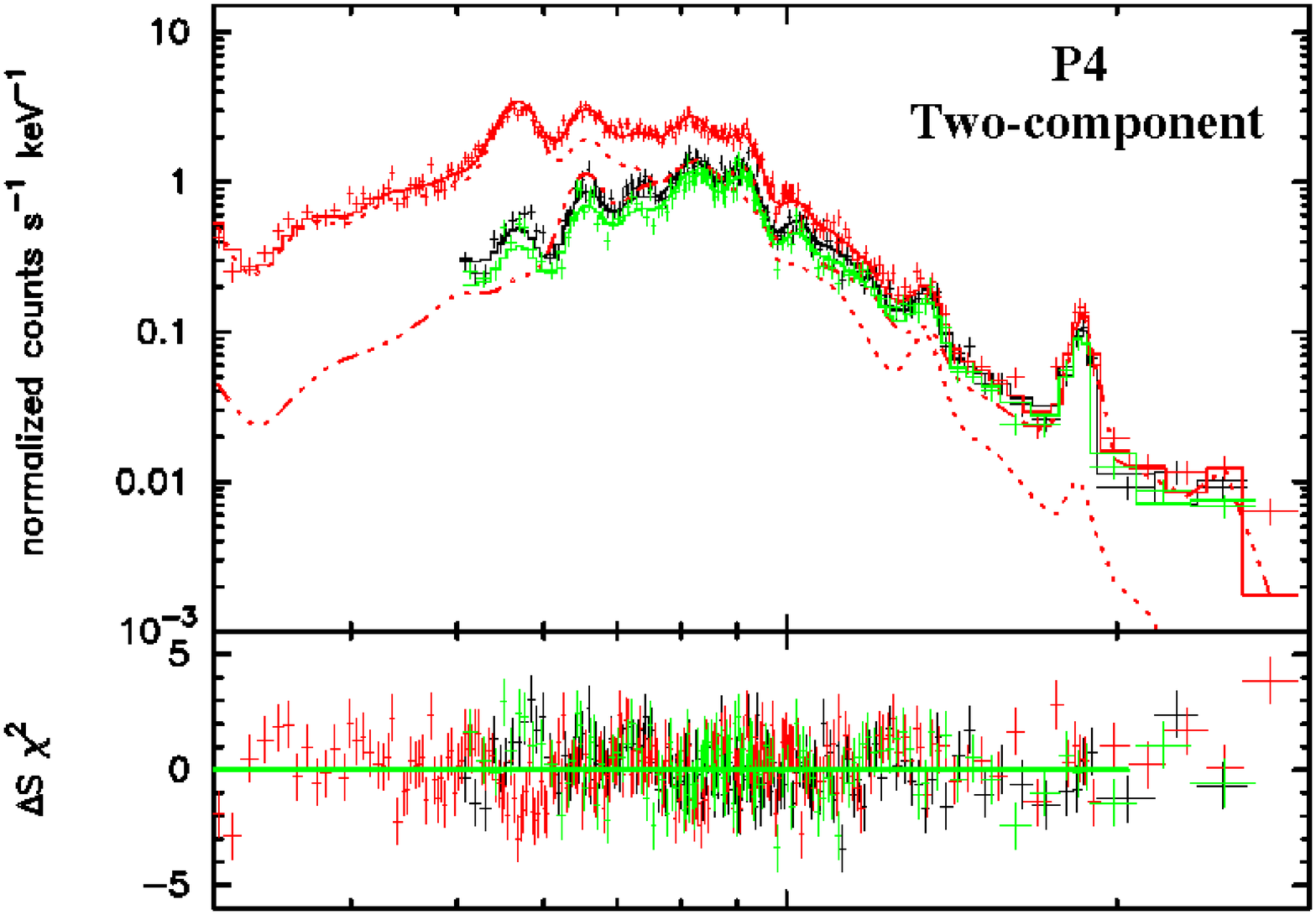}
   \includegraphics[width=70mm]{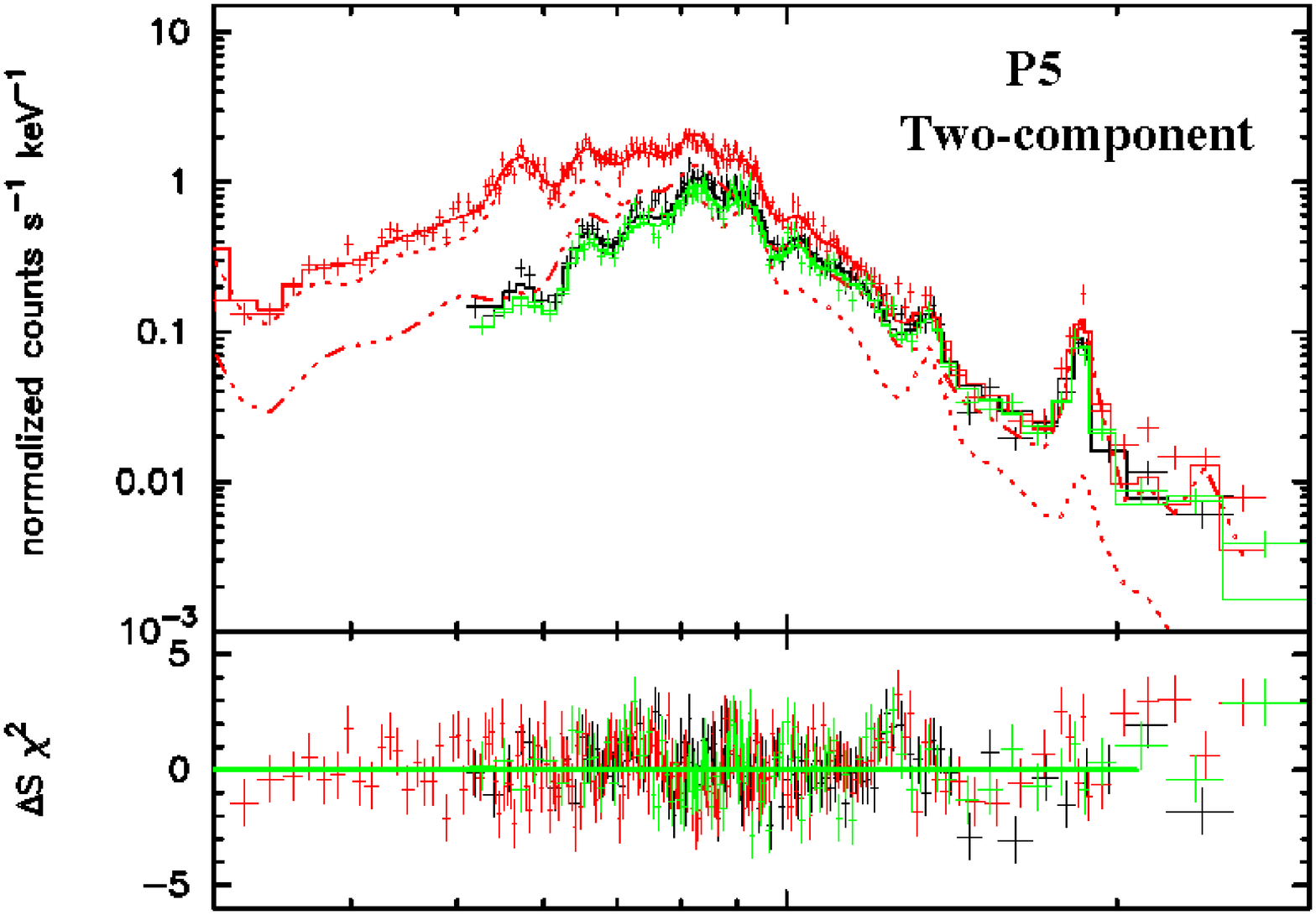}
   \includegraphics[width=70mm]{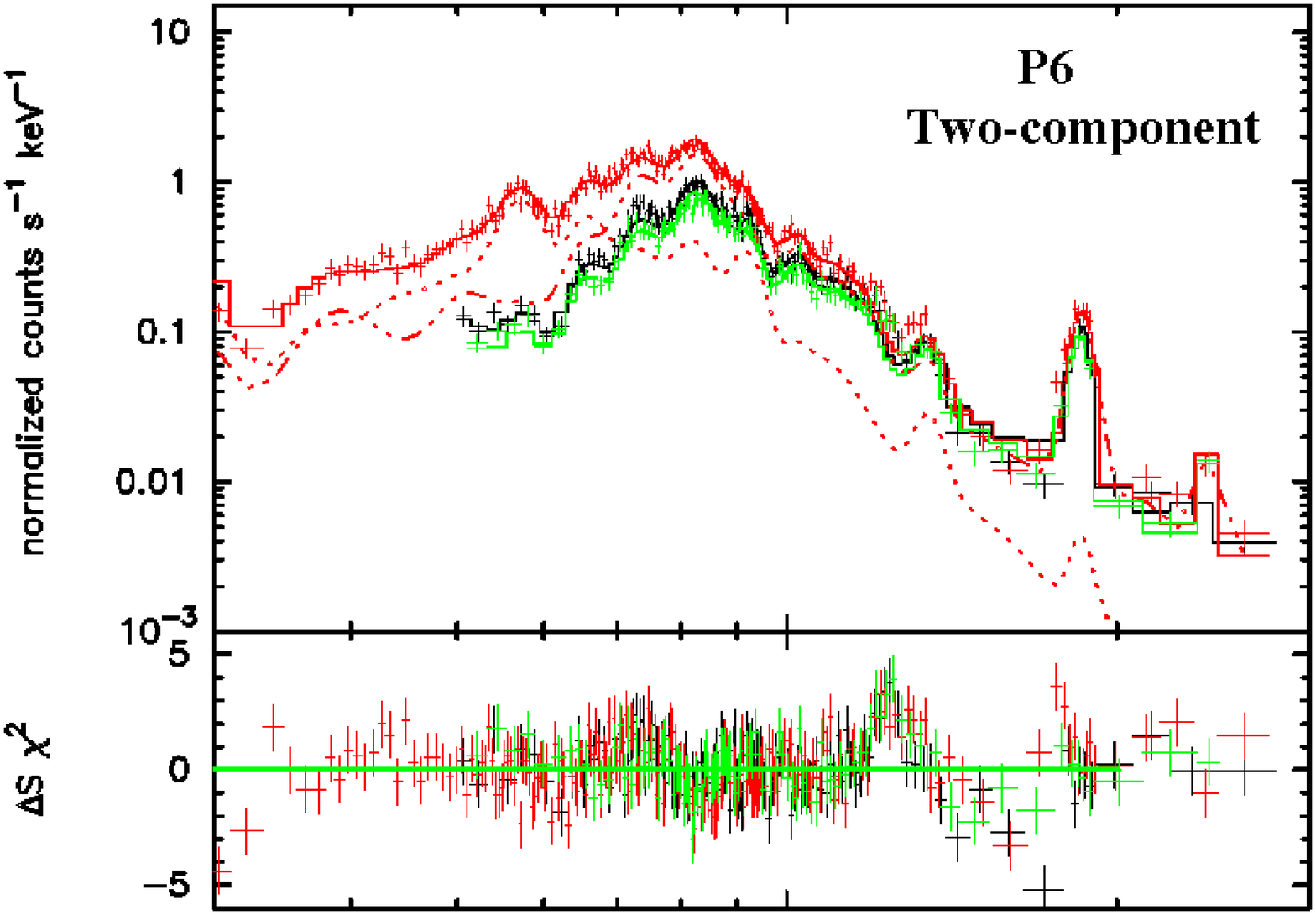}
   \includegraphics[width=70mm]{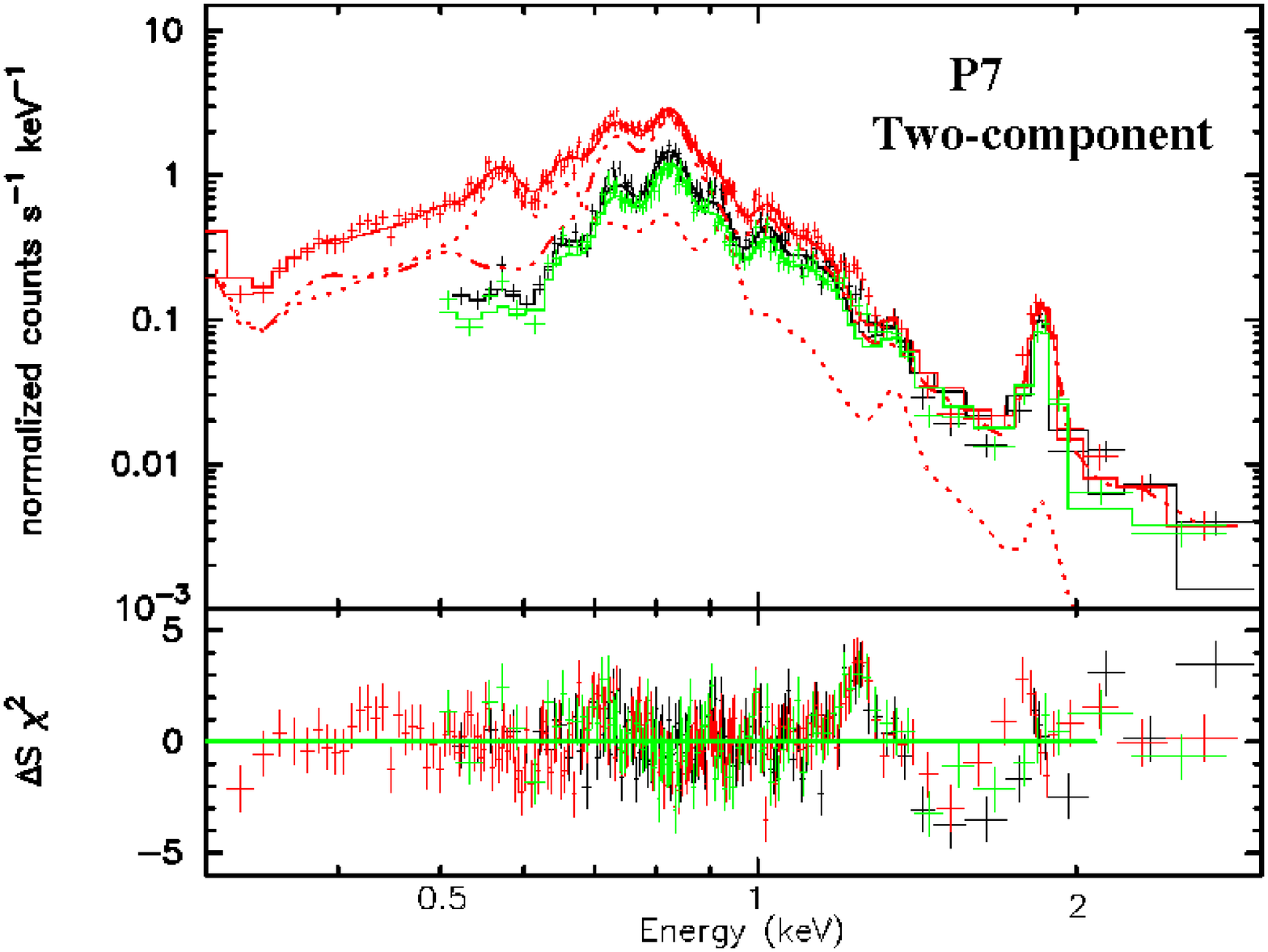}
   \includegraphics[width=70mm]{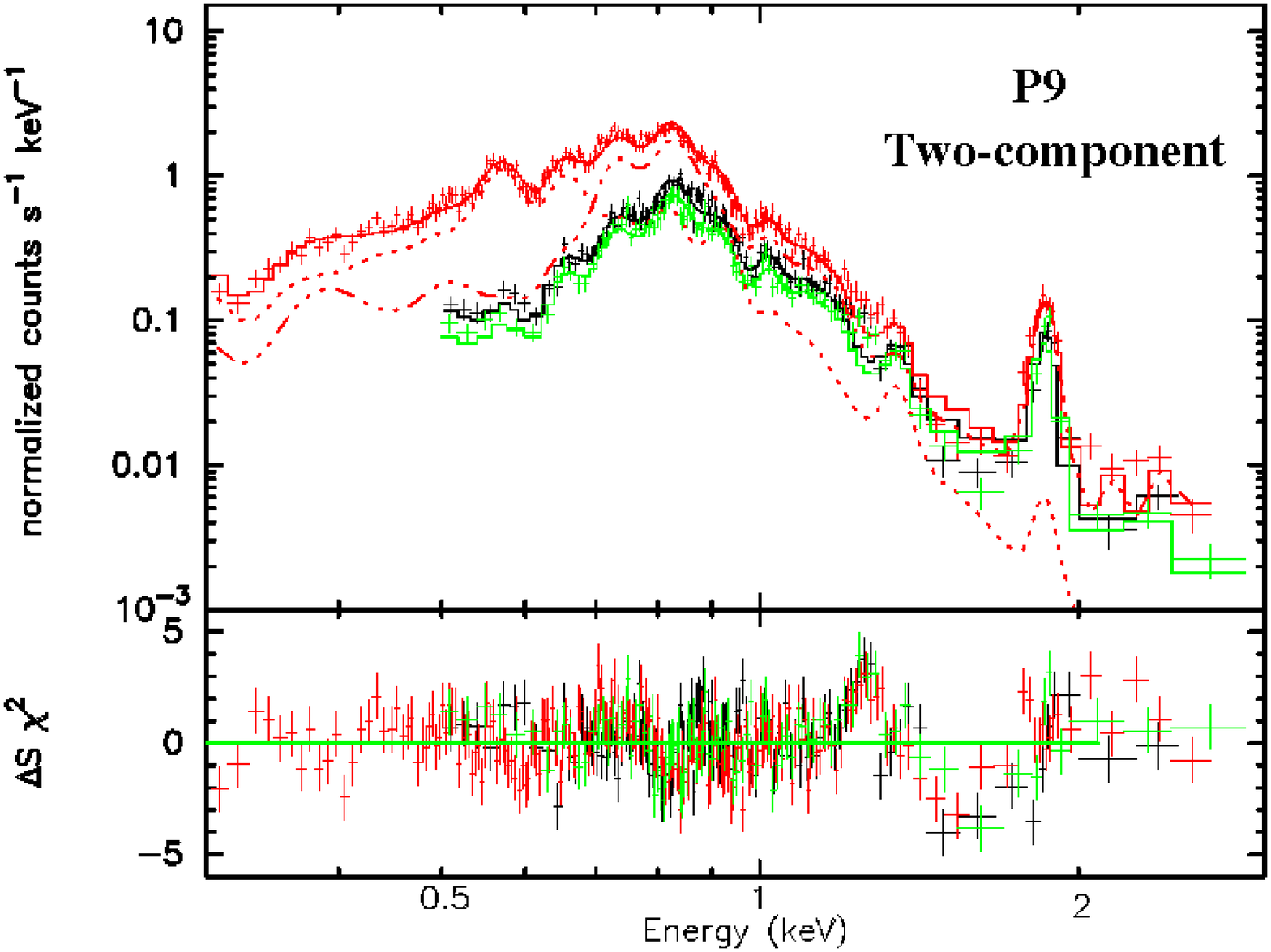}
 \caption{X-ray spectra extracted from the red regions in Figure~\ref{fig:xis1_image}. The best-fit curves of two-component VNEI models are shown in solid line. The contributions of each component are shown in dotted lines.}  
\label{fig:spec} 
\end{center}
\end{figure*}

We tried to apply this model by freeing all parameters, but since this did not give as reasonable values, we decided to fix certain values.
The \kTe, $\tau$, and EM are free parameters in both components.  $N_\mathrm{H}$ is also a free parameter but shares same value between two components. 
Since abundances of the swept-up matter region is well known, we set the abundances of low-\kTe -component to those values determined in the NE region by Uchida et al. (2006). The abundances of O (=C=N), Ne, Mg, Si (=S), and Fe (=Ni) for the high-\kTe -component are set as free parameters.
The fit improved significantly ($>99\%$ based on F-test probability) in non-rim region ($-70\arcmin<R<65\arcmin$) by adding another component (see figure 3). On the other hand, the NE rim ($R<70\arcmin$) and the SW rim ($R>65\arcmin$) did not show notable improvement on the fit. Furthermore, parameters of the high-\kTe-component had large uncertainties in rim regions. Therefore, we employed two-\kTe-component model in the non-rim regions ($-70\arcmin<R<65\arcmin$), while one-\kTe-component model in the rim regions.
Figure~\ref{fig:spec} shows spectra extracted from red regions shown in figure~\ref{fig:xis1_image}. 
The top 2 panels are fitted with one-component model (NE3 and P10) while the bottom 8 panels are fitted with two-component model.
The best-fit parameters are summarized in table~\ref{param}. 
The residuals between the data and the model in bottom 4 panels (P5, 6, 7, 9) of figure~\ref{fig:spec} show some structures around 1.2keV. 
We should keep in mind that the two-\kTe~VNEI model still did not give us acceptable fits from the statistical point of view, which suggests that our model is too simple.

\begin{figure*}
\begin{center}

\includegraphics[width=7cm]{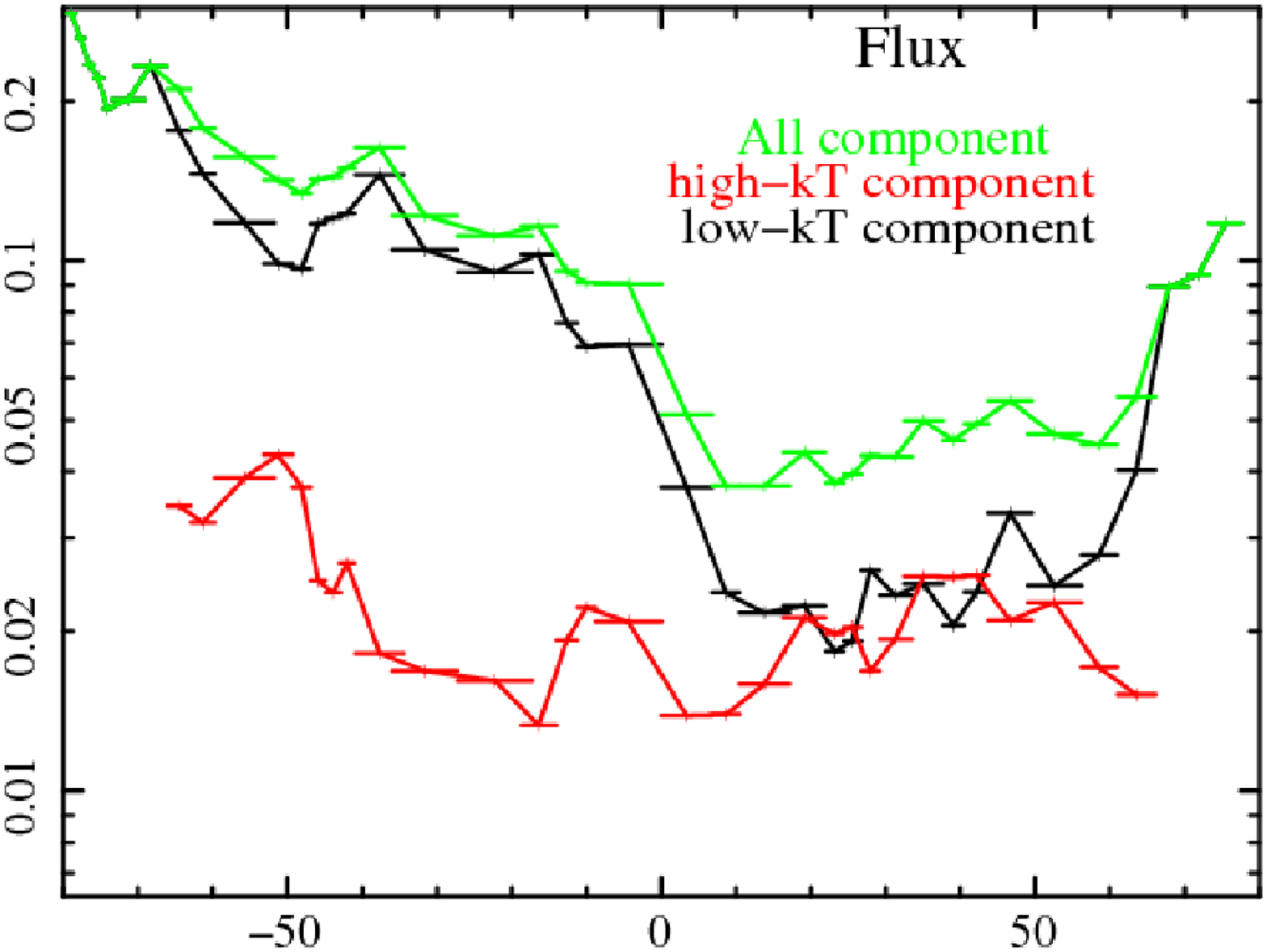} 
\includegraphics[width=7cm]{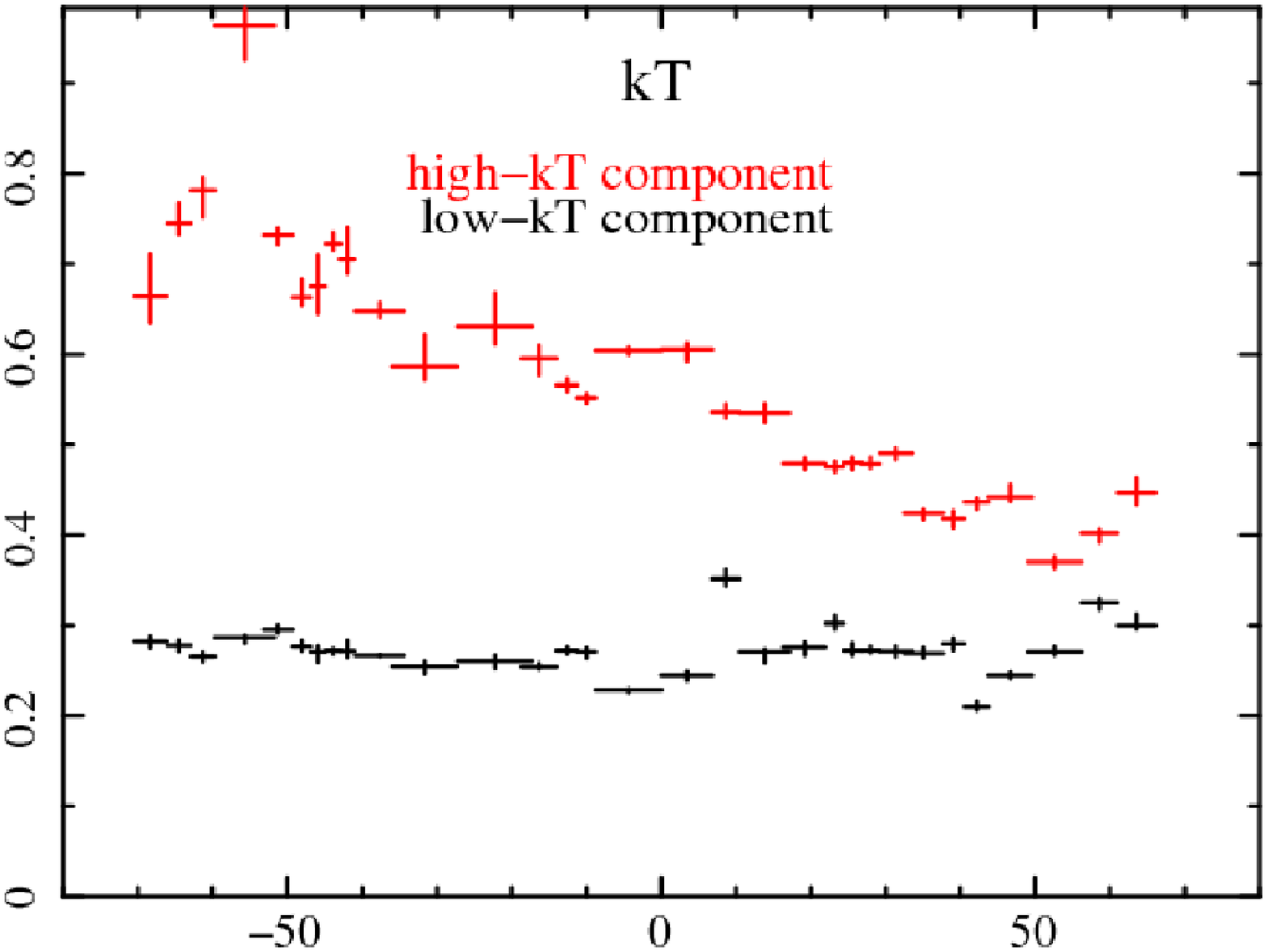} 
\includegraphics[width=7cm]{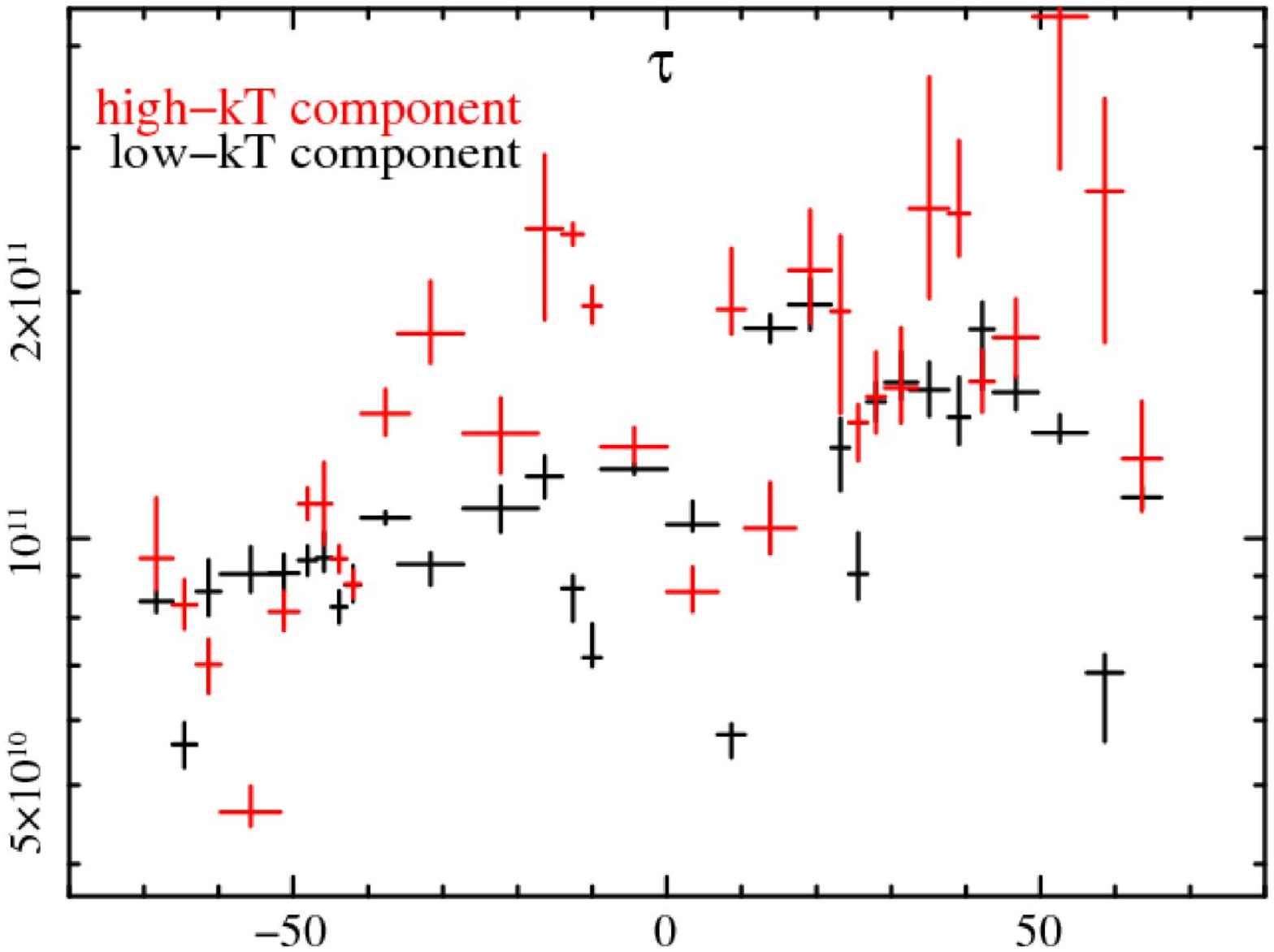} 
\includegraphics[width=7cm]{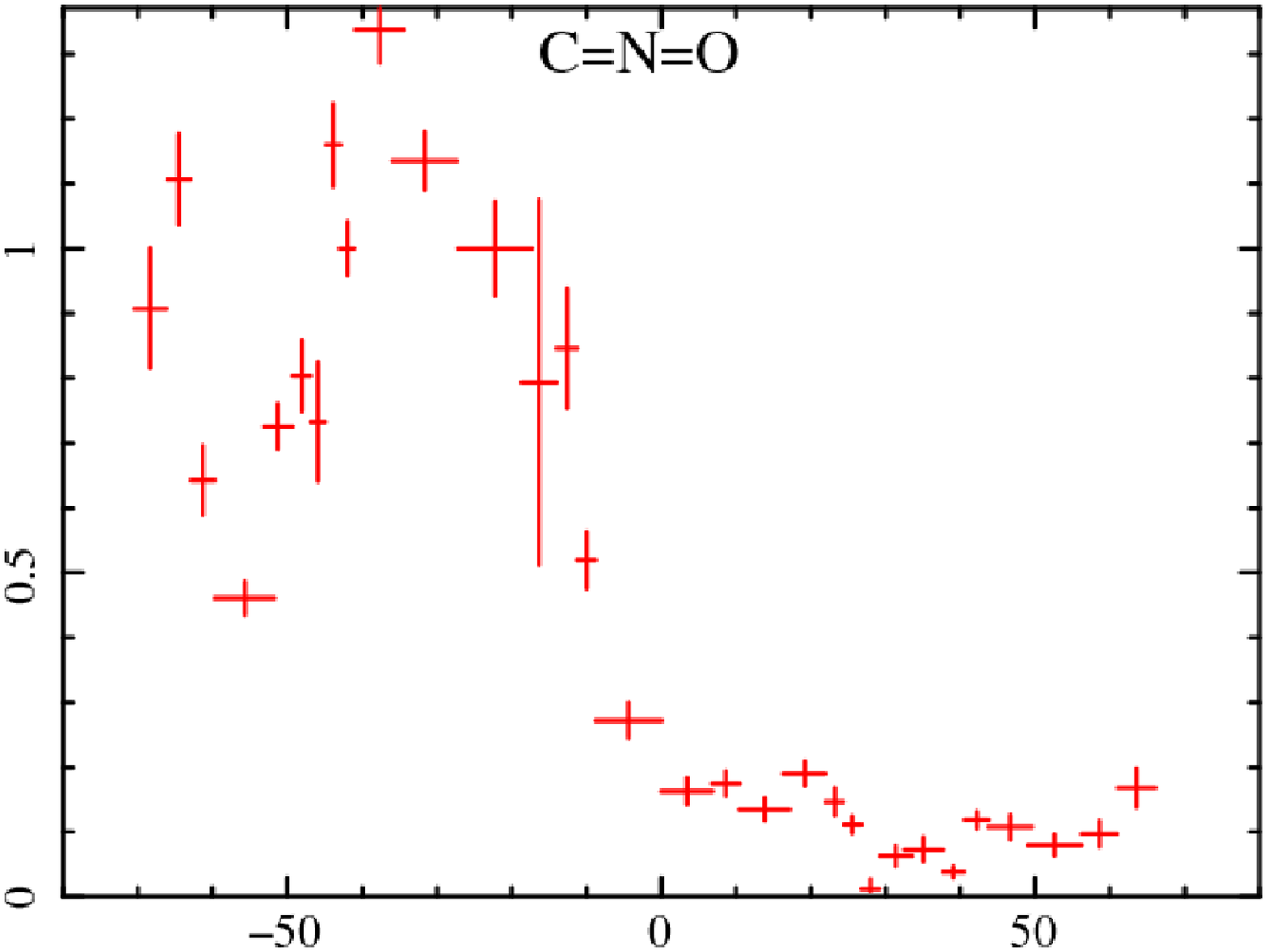} 
\includegraphics[width=7cm]{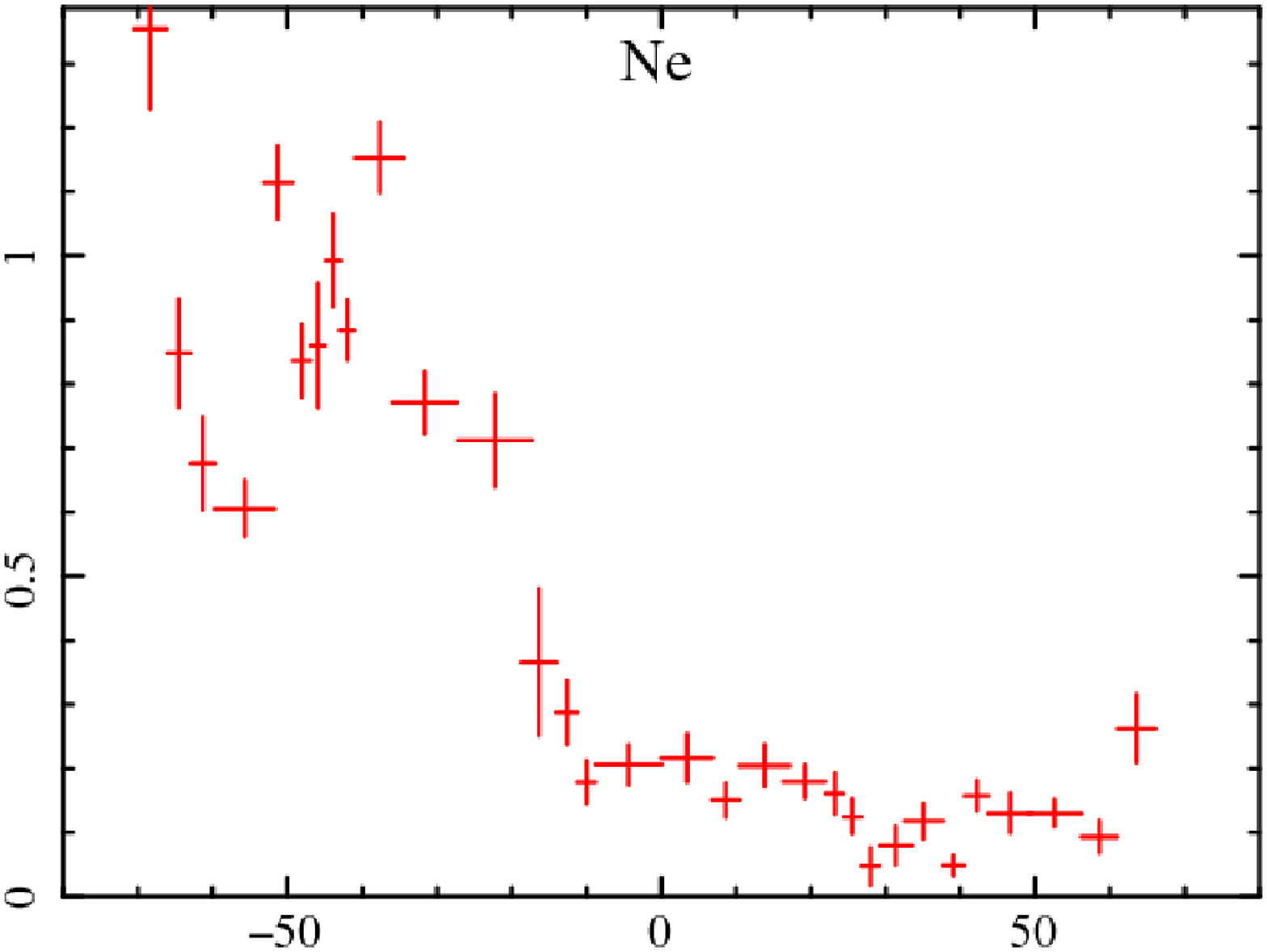} 
\includegraphics[width=7cm]{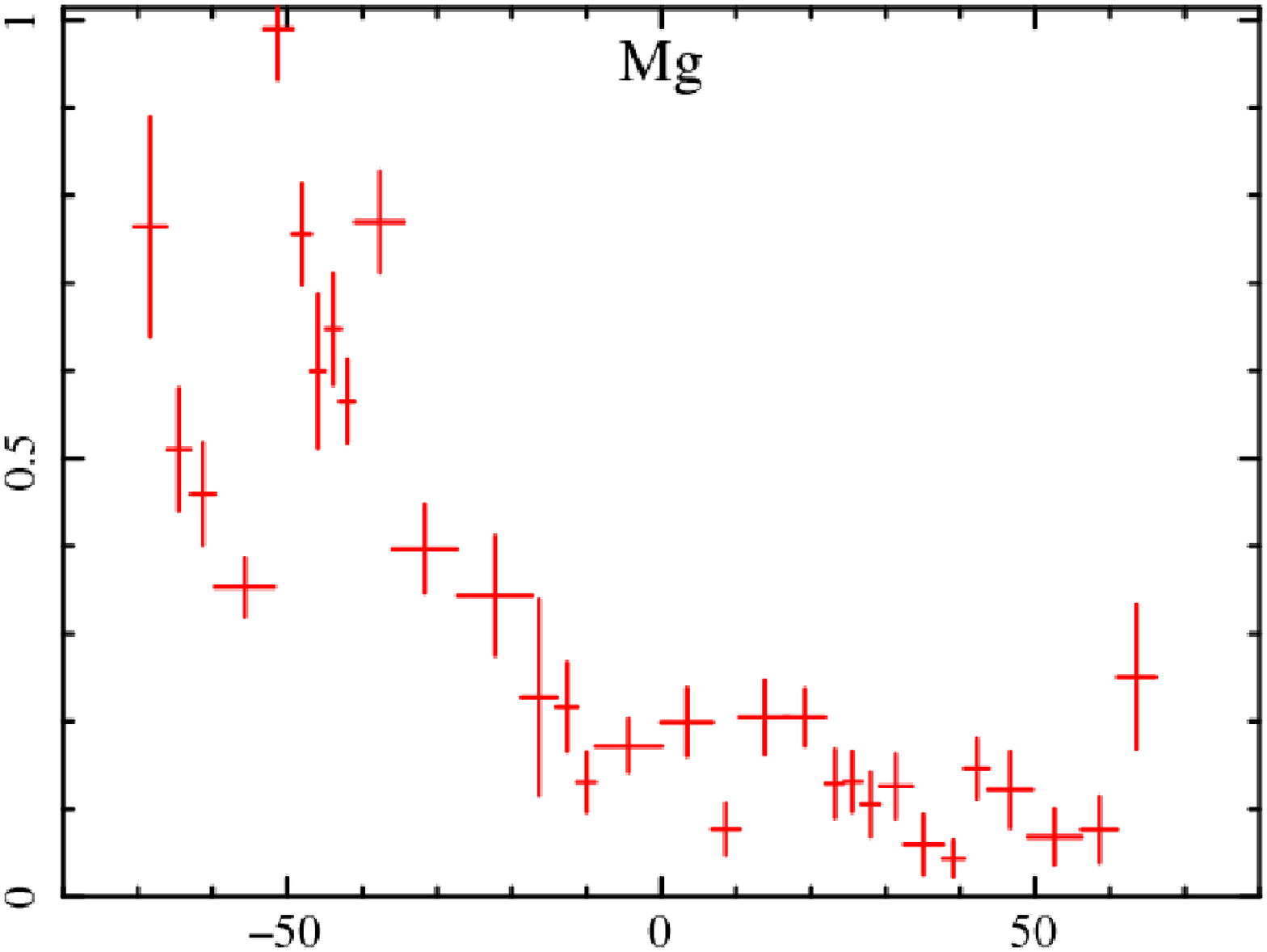} 
\includegraphics[width=7cm]{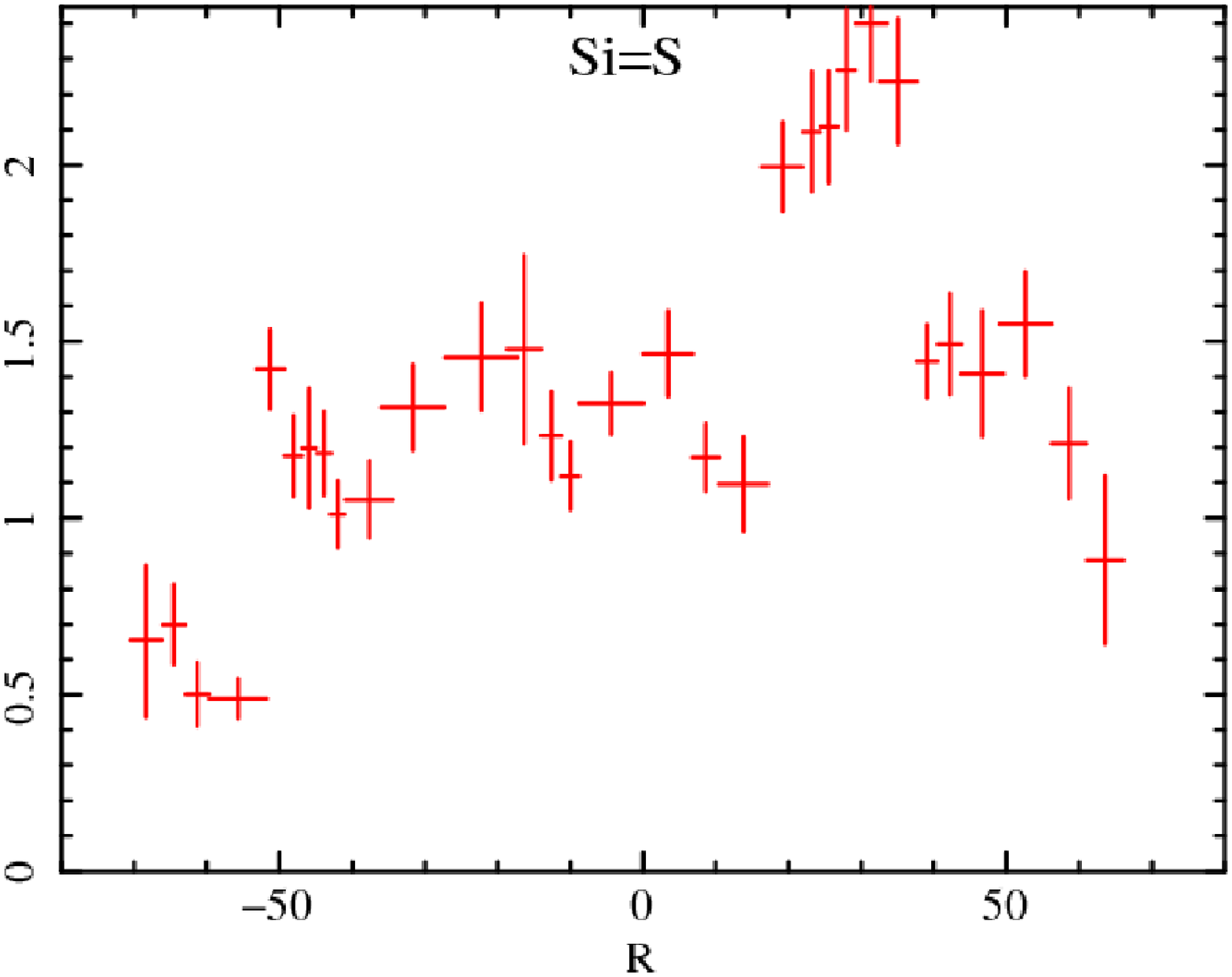} 
\includegraphics[width=7cm]{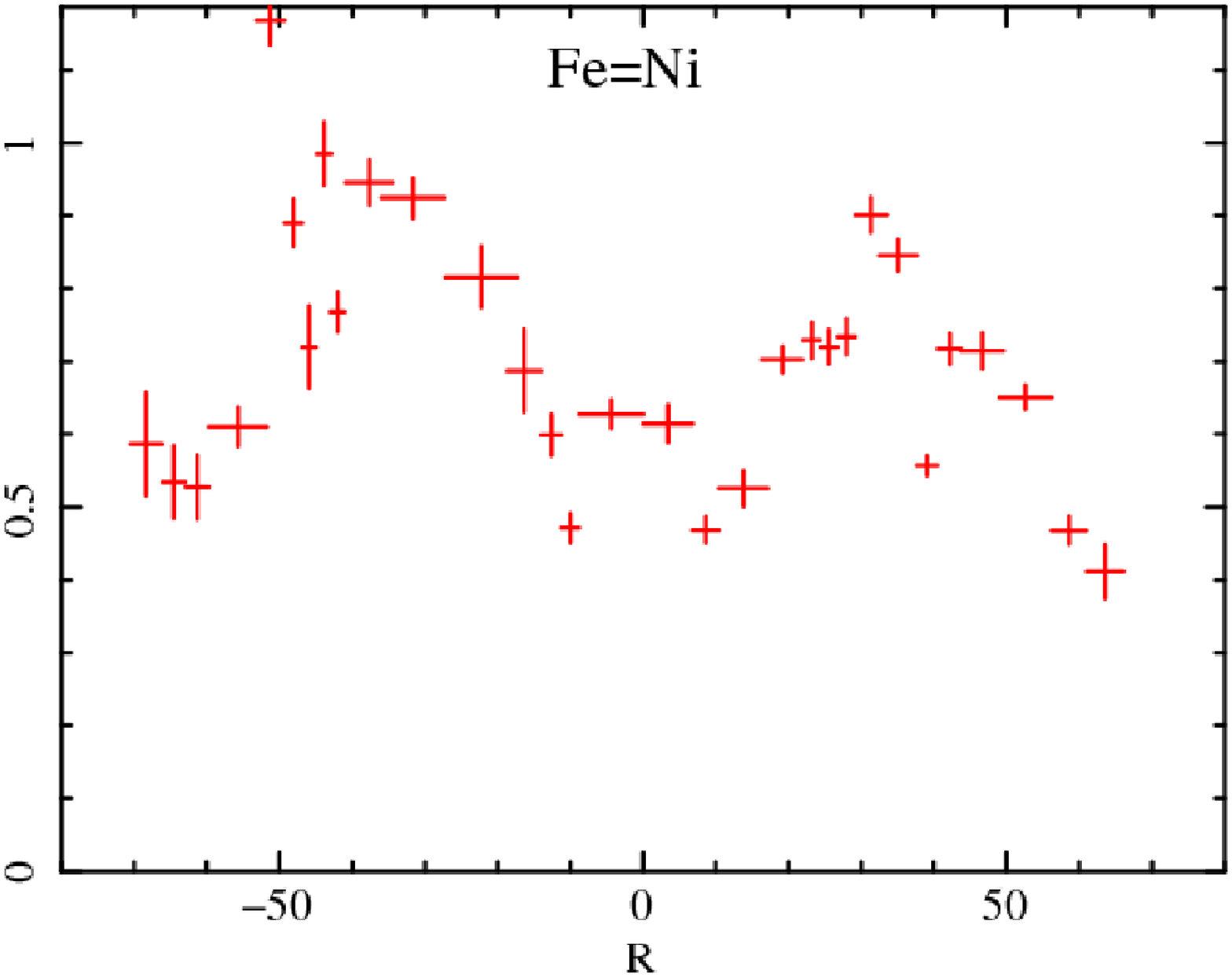} 
\caption{The $kT_e$, $\tau$ values and abundance of elements as a function of $R$.  The black marks show the parameters of low-\kTe~for comparison.}
\label{fig:abund} 
\end{center}
\end{figure*}

Figure ~\ref{fig:abund} shows $kT_e$, $\tau$ and metal abundance as a function of $R$. 
We found that the high-\kTe -component shows distinctly higher temperature and metal abundance compared to low-\kTe -component. 
The temperature of the low-\kTe -component stays almost constant ($\sim$ 0.3 keV) while the temperature of the high-\kTe -component has a peak value of 0.8 keV around $R=-60\arcmin$ and decrease towards the SW.  We also found asymmetry in metal abundance of high-\kTe -component. The abundance of O[=C=N], Ne and Mg are higher in the NE part ($R<0\arcmin$) while  Si is in the SW ($R>0\arcmin$).  
The flux of each component is shown in top left panel of figure ~\ref{fig:abund}. The flux of the low-\kTe-component is dominant in the NE part and it also shows shell brightening in both NE and SW rims, therefore, we confirm that low-\kTe-component surrounds the Loop and the high-\kTe-component fill its interior .
All of these facts support our assumption where low-\kTe -component originates from the swept-up matter and the high-\kTe -component originates from the ejecta.

\section{Discussion and Conclusion}
We observed the Cygnus Loop from the NE to the SW with Suzaku in ten pointings. 
Our FOV  covers more to the north of previous observation carried out by XMM-Newton and Suzaku.
Dividing the entire FOV into 45 rectangular regions, we extracted spectra from all the regions, and performed spectral analysis to them.
For spectral fit, we employed  models used by the previous observation (Katsuda et al. 2008; Tsunemi et al. 2007) and obtained similar results.
The one-component VNEI model showed fairly good fit in the rim of the Loop but not  in the non-rim part of the Loop.  The fit of non-rim part improved significantly  by applying two-component VNEI model.
Judging from the abundanes and the flux distribution of each component, the high-$kT_e$ component must be the ejecta, while the low-$kT_e$ component comes from the swept-up matter. 

\subsection{Swept-up Matter Distribution}
The temperature for the swept-up matter component is significantly
lower than that for the ejecta component. This temperature is
similar to that obtained for the rim of the Loop, where we expect no
contamination of the ejecta (e.g., Miyata et al. 2007).
Therefore, we believe that we surely separated the X-ray emission of the
ejecta inside the Loop from that of the surrounding matter. The EM
distribution of the swept-up matter is inhomogeneous
in our FOV, as shown in top right panel of figure~\ref{fig:em}. The shell of the swept-up matter seems to be thin around $10\arcmin<R<60\arcmin$ relative to that in the NE part ($R<0\arcmin$). 
The flux of swept-up matter also shows this tendency; its flux in the SW part is about a third of the NE part. 
Such trend is also reported by XMM-Newton (Tsunemi et al. 2007) and previous Suzaku observations (Katsuda et al. 2008).
Tsunemi et al. (2007) divided XMM-Newton observation into north path and south path and found that this thin shell region to be $5^{\prime}$ in south path and $20^{\prime}$ in north path.  They estimated this thin shell region to have diameter of $1^{\circ}$ and centering $(20^{h}49^{m}11^{s}, 31^{\circ}50^{\prime}20^{\prime\prime})$.
Our observation path goes though right in middle of this region and our result was consistent with their prediction.
Since the EM of this region is also about $\frac{1}{4} - \frac{1}{3}$ of the NE part,
and the EM of ejecta is stronger compared to the NE part,
there might be a blowout in the direction of our line of sight, just like a blowout in the south of the Loop (Uchida et al. 2008) and we are only seeing the one side of the shell.
We detected no ejecta emission from the NE rim ($R\leq -70\arcmin$) and the SW rim ($R \geq 65\arcmin$) of the Loop, which is about 15$\%$ of our FOV. 
The ejecta occupy the major part of the inner side of the Loop while the swept-up matter surrounds it. This structure is identical to those in previous results (Miyata et al. 1998; Katsuda \& Tsunemi 2008).

\subsection{Metal distribution in Ejecta }  
\begin{figure*}
\begin{center}

\includegraphics[height=6cm]{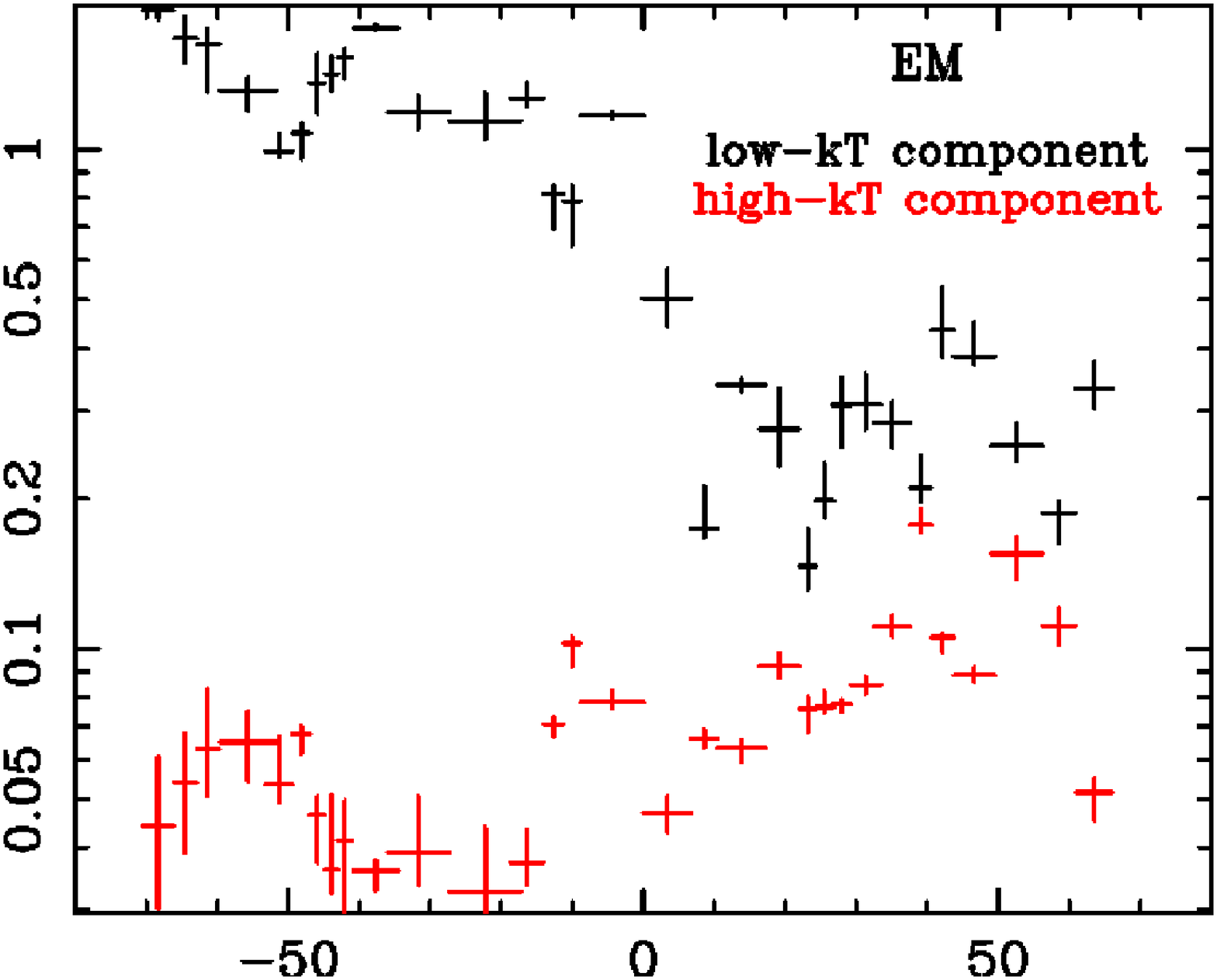} 
\includegraphics[height=6cm]{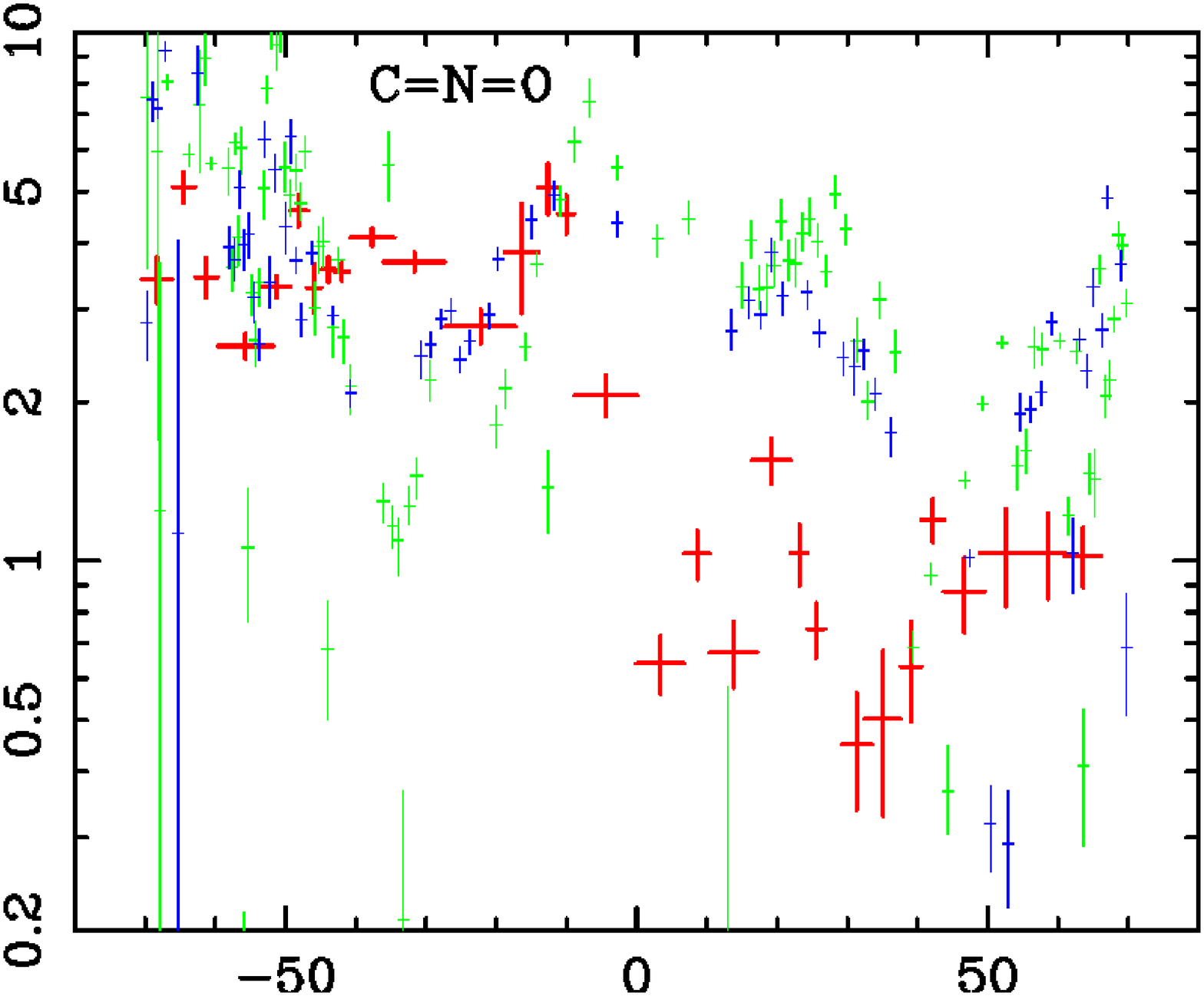}  
\includegraphics[height=6cm]{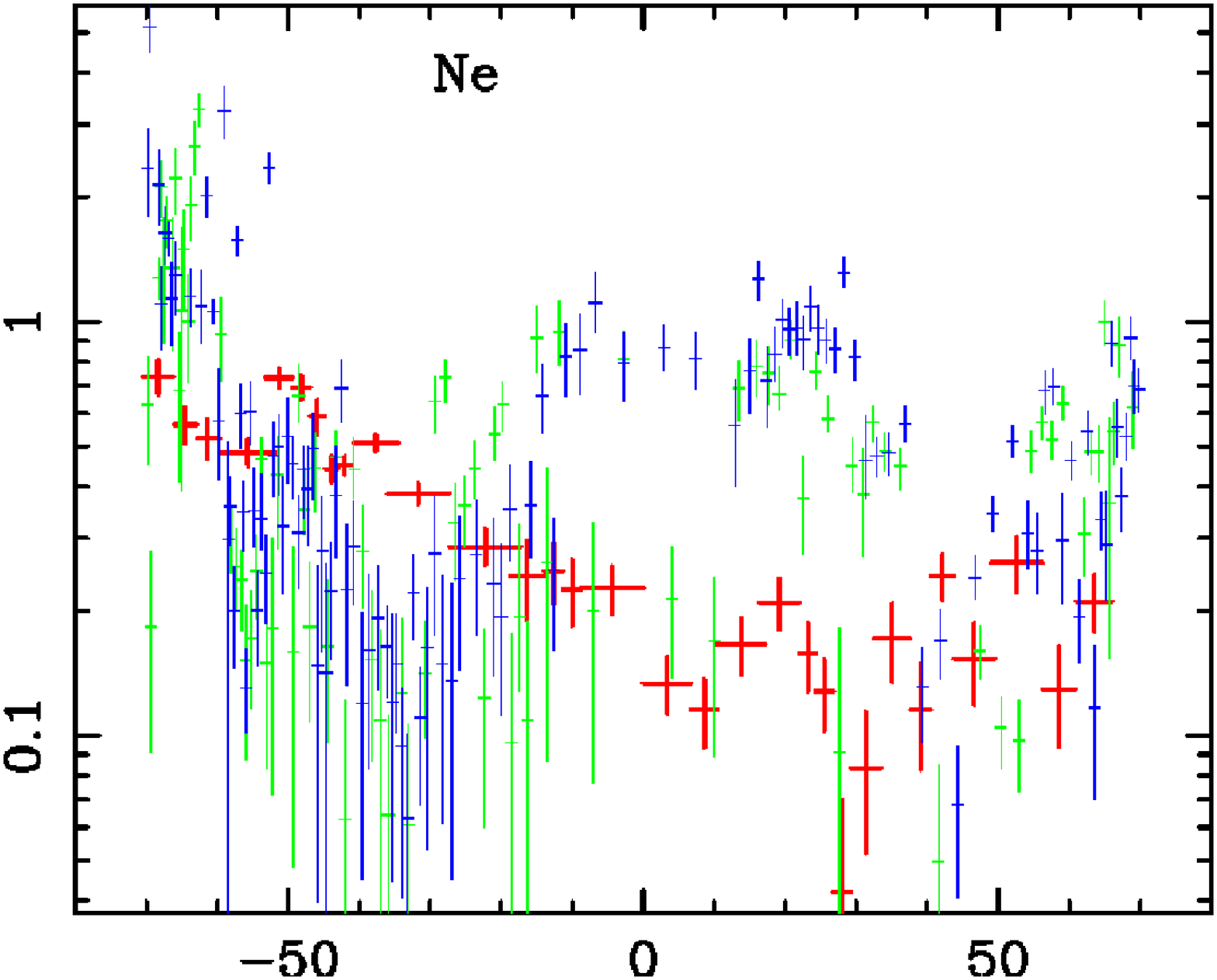} 
\includegraphics[height=6cm]{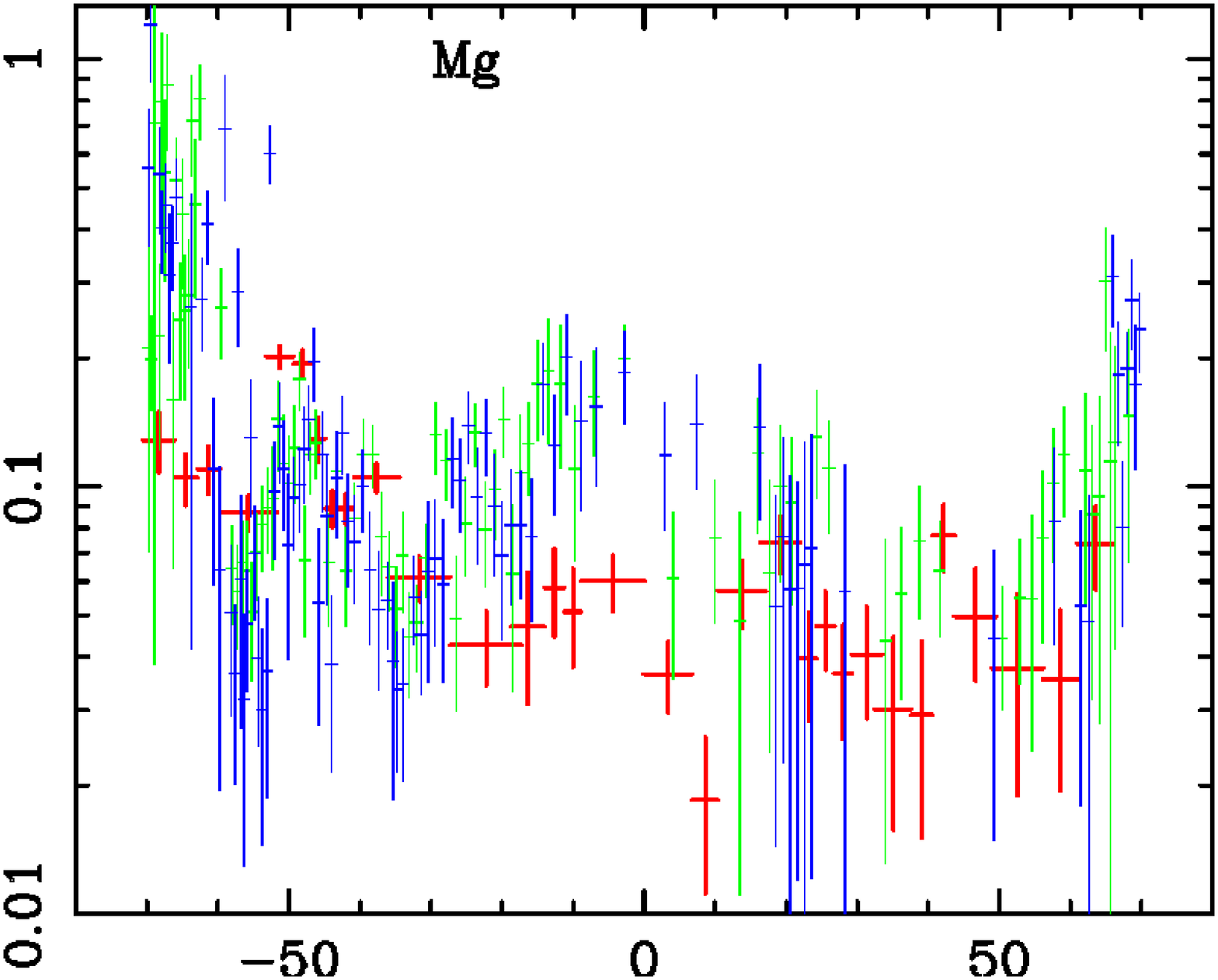} 
\includegraphics[height=6cm]{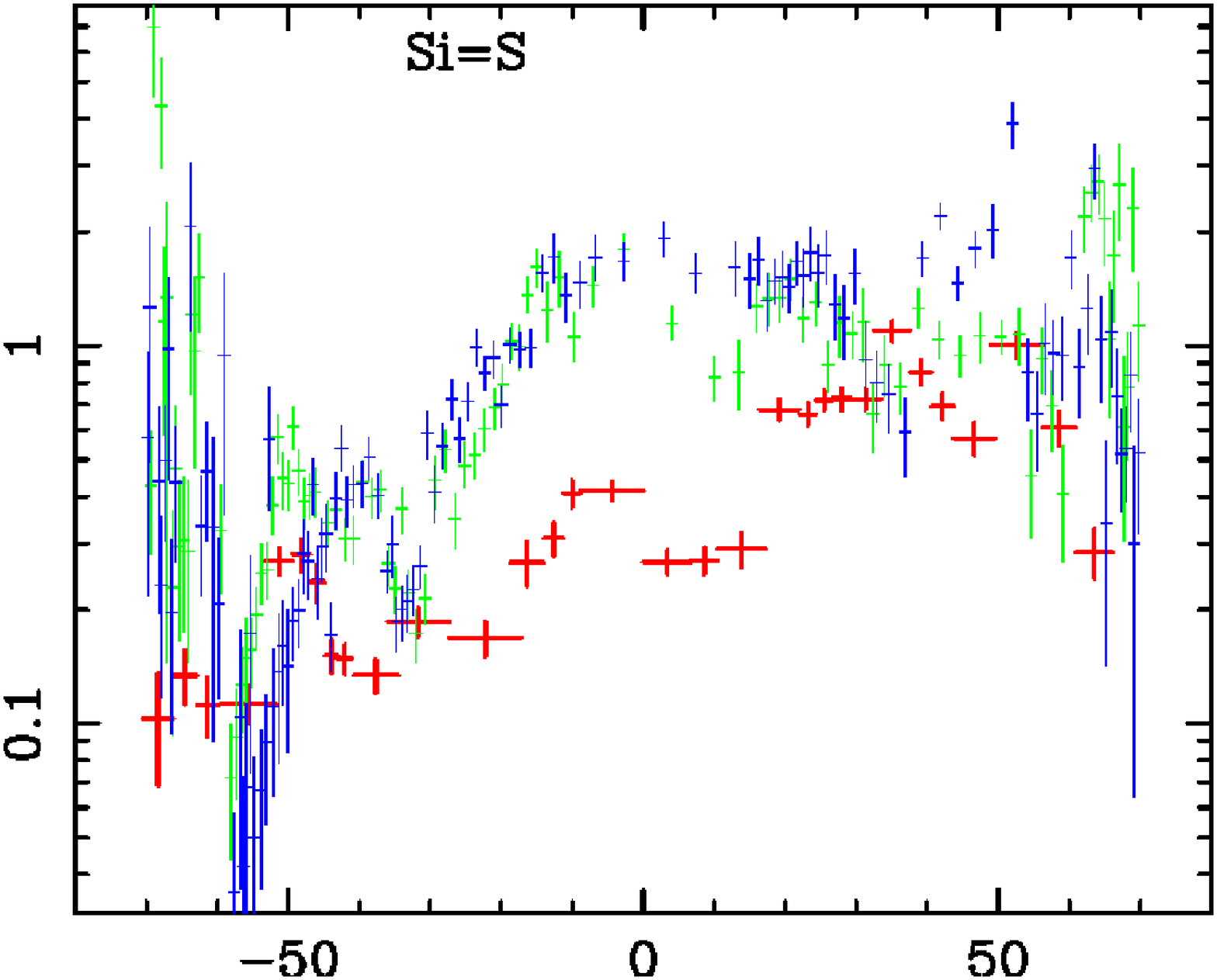} 
\includegraphics[height=6cm]{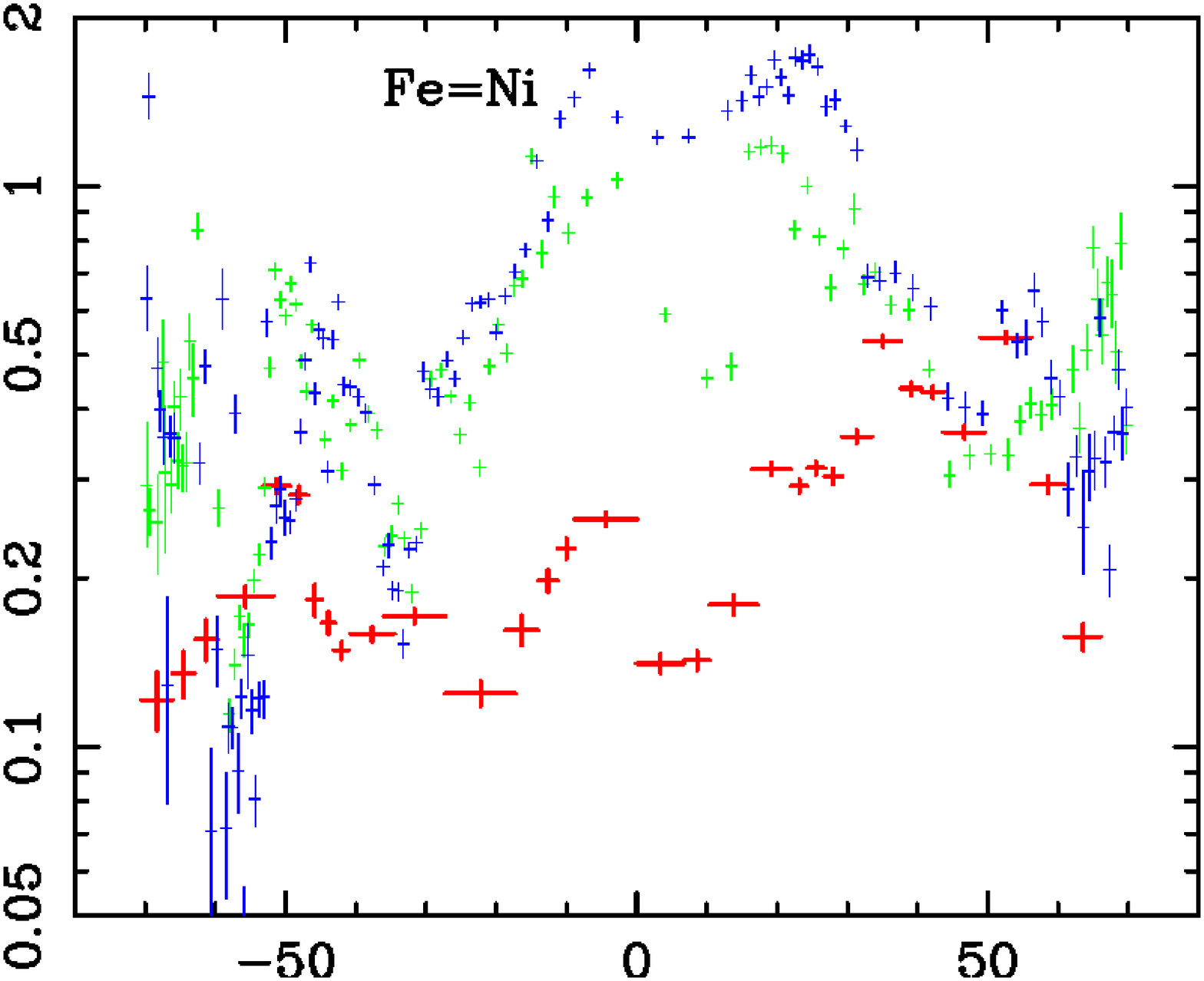}

\caption{The top left panel shows the EM ($\int n_en_H dl$) in $10^{19}$cm$^{-5}$. The other panels are EM of elements ($\int n_en_X dl$) in $10^{14}$cm$^{-5}$. The results from our Suzaku FOV is shown in the red mark while the green and the blue mark shows the results from XMM-Newton north path and south path (Tsunemi et al. 2007). }
\label{fig:em} 
\end{center}
\end{figure*}

 We have calculated the EM of various heavy elements in the ejecta such as O[=C=N], Ne, Mg, Si, and Fe [=Ni]. Figure~\ref{fig:em} shows the EM distribution for these elements as a function of $R$. 
The red marks show the results from our FOV while the green and the blue marks show the results from XMM-Newton north path and south path.  The comparison of these results show some similarity near the both rim of the Loop ($R<-60\arcmin, 40\arcmin<R$), but shows discrepancy in the center portion of the Loop. This is especially notable in Si and Fe.  Since our FOV is further north from the geometric center of the Loop compared to XMM-Newton observation, these discrepancies  suggests onion layer structures at a SN time of explosion.
For Si and Fe, some structures can be seen in both Suzaku and XMM-Newton observations, such as the bump around $R=-50\arcmin$ and the drop at $0\arcmin<R<20\arcmin$.

From the result from our FOV, the NE part and the SW part have different characteristic of metal distributions. We calculated the mass ratio in the NE part and those in the SW part from Figure \ref{fig:em}. 
  They are O$\sim 4.7$, Ne$\sim 3.2$, Mg$\sim 2.3$, Si$\sim 0.36$, Fe$\sim 0.62$ assuming unity filling factor.
The O-Ne-Mg group is heavily distributed in the NE part of the Loop by factor of $2-5$. In contrast to the O-Ne-Mg group, the Si-Fe group is distributed more in the SW part of the Loop by factor of $2-3$.
These asymmetries can also be seen in all of the abundance distribution (figure \ref{fig:abund}) except for Fe.
However EM of the elements is the clearer representation of the amounts of elements, therefore we believed that Fe is also distributed asymmetrically more to the SW.
This asymmetry is quantitatively consistent from previous observation (Katsuda et al. 2008).
A natural explanation for the asymmetry is an asymmetry at the time of the SN explosion of the Cygnus Loop.
Recent theoretical models describe asymmetric supernova explosion resulting from hydrodynamic instability (Burrows et al. 2007).  

\begin{figure*}
\begin{center}
   \includegraphics[height=8cm]{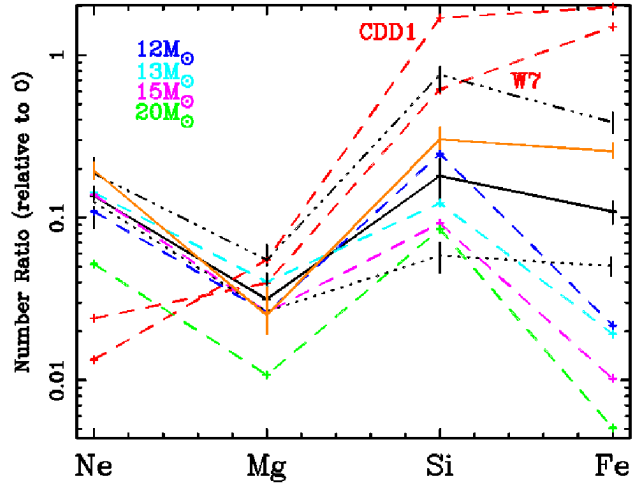} 
\caption{Number ratios of Ne, Mg, Si, Fe, relative to O of the high-\kTe -component for the entire FOV (black line). Black dotted line is a number ratio for NE part while black dot-dash line is for SW part of the Loop.
The orange line is a number ratio for all of the 2 Suzaku observations and XMM-Newton observations (Katsuda et al. 2008; Tsunemi et al. 2007). 
Dotted red lines represent the CDD1 and W7 Type Ia SN models of Iwamoto et al. (1999). Dotted blue, light blue, magenta, and green lines represent core-collapse models with progenitor masses of 12, 13, 15, 20 M$_\odot$ respectively (Woosley \& Weaver 1995).}
\label{fig:mass} 
\end{center}
\end{figure*}

We calculated the mass of ejecta to be 21M$_\odot$ from EM.  
We estimated the plasma depth to be 26pc, unity filling factor and $n_e=1.2n_H$, although the fossil ejecta might be deficient in hydrogen. If this is the case, the total mass of the ejecta reduces to $\sim$12M$_\odot$
However this value strongly depends on the plasma structure of the ejecta.  Therefore we used SN explosion model to calculate the mass.  
In order to compare our data with the SN explosion models, we calculated the ratios of elements relative to O. Figure~\ref{fig:mass} shows the number ratios of Ne, Mg, Si and Fe relative to O of the ejecta component. We plotted the core-collapse model (Woosley \& Weaver 1995) for the various progenitor mass and Type Ia supernova model (Iwamoto et al. 1999) for comparison. Since the distribution characteristic of elements showed large difference between the NE and the SW, we also plotted the number ratios calculated from each part.
The orange line represents the number ratios calculated from all of observations including XMM-Newton and previous Suzaku observations (Tsunemi et al. 2007, Katsuda et al. 2008).
The number ratios from the SW part shows fairly good agreement with Type Ia models in Mg and Si but that the number ratios of Ne and Fe show large inconsistency from the model.
On the contrary, the number ratios of Ne, Mg, Si, from NE are in agreement with 15M$_\odot$  model, but Fe is about 4 times higher in our data.
Because of the asymmetric distributions of ejecta, the number ratios of heavy elements give very different values by choosing which part of the Loop we use. 
The number ratios calculated from our FOV and all FOV (2 Suzaku and 1 XMM-Newton) are both in good agreement with that of 12M$_\odot$ model.
However, just like any other data, the number ratio of Fe shows disagreement.
It is possible that the Fe is overabundant because all FOV are across the center of the Loop, where we expect to be Fe rich region.



\section*{Acknowledgements}

This work is partly supported by a Grant-in-Aid for Scientific
Research by the Ministry of Education, Culture, Sports, Science and Technology (16002004).  This study is also carried out as part of the 21st Century COE Program, \lq{\it Towards a new basic science: depth and synthesis}\rq.  H.U. and S.K. are supported by JSPS Research Fellowship for Young Scientists.
We also like to thank Hiroko Kosugi for careful reading of this manuscript.
 \newpage

\begin{table*}
 \begin{center}
 \caption{Information of observations of the Cygnus Loop and Lockman Hole}
  \begin{tabular}{lccc}
\hline
Obs. ID &Coordinate (RA, DEC) &Obs. Date& Effective Exposure\\
\hline
\multicolumn{4}{c}{Cygnus Loop}\\
500022010 (NE3) & 313.746, 32.188 & 2005.11.29 & 12.2\,ks\\
501012010 (P1)  & 313.510, 31.975 & 2007.11.13 &  9.8\,ks\\
501013010 (P2)  & 313.265, 31.779 & 2007.11.14 & 16.4\,ks\\
501014010 (P3)  & 313.032, 31.574 & 2007.11.14 &  7.5\,ks\\
501015010 (P4)  & 312.799, 31.369 & 2007.11.14 & 18.3\,ks\\
501016010 (P5)  & 312.547, 31.180 & 2007.11.15 & 19.3\,ks\\
501017010 (P6)  & 312.297, 30.991 & 2007.11.11 & 28.7\,ks\\
501018010 (P7)  & 312.078, 30.776 & 2007.11.12 & 21.0\,ks\\
501019010 (P9)  & 311.809, 30.603 & 2007.11.12 & 16.2\,ks\\
501020010 (P10) & 311.566, 30.407 & 2007.11.13 & 14.6\,ks\\
\hline
\multicolumn{4}{c}{Lockman Hole}\\
100046010 (for NE3) & 163.4063, 57.6108 & 2005.11.14 &49.3\,ks\\
102018010 & 162.9257, 57.2581 & 2007.05.03 & 68.9\,ks\\
\hline
\label{obs}
  \end{tabular}
 \end{center}
\end{table*}
\begin{sidewaystable}
\caption{Best-fit parameter for example spectra. NE and SW (P10) are fitted with one-component VNEI model while the others are fitted with two-component VNEI model }
\begin{center}
\tabcolsep4pt
\begin{tabular}{lcccccccccc}
\hline\hline
Elements & NE3 &P1 & P2 &P3& P4 & P5 & P6 & P7&P9 & SW(P10) \\
\hline
\multicolumn{11}{c} {Free abundance component} \\
$n_H [\times 10^{20}\mathrm{cm}^{-2}]$ \dotfill
 &$3.8\pm0.7$&$4.1\pm0.8$&$3.3\pm0.5$&$4.5\pm0.5$&$3.1\pm0.6$
 &$3.2\pm0.5$&$3.2\pm0.5$&$3.6\pm0.4$&$3.5\pm0.5$&$3.2\pm0.6$ \\

\kTe [keV] \dotfill& 0.28 $^{+0.03}_{-0.04} $& 0.74 $^{+0.02}_{-0.01}$ & 0.66 $\pm$0.01 &0.63$\pm0.02$& 0.55$\pm$0.01
		   & 0.53$\pm$0.02 &0.48$\pm$ 0.01 & 0.41 $\pm$0.01 &0.40 $\pm0.01$&  0.32 $\pm$0.01  \\
C \dotfill	&$0.3\pm 0.1$&$=$O&$=$O&$=$O&$=$O&$=$O&$=$O&$=$O&$=$O&0.38$\pm0.08$ \\
	
N	\dotfill&$0.13^{+0.1}_{-0.05}$&$=$O&$=$O&$=$O&$=$O&$=$O&$=$O&$=$O&$=$O&$0.15^{+0.08}_{-0.05}$ \\
		   
O \dotfill& 0.10 $^{+0.1}_{-0.04}$&1.10 $\pm 0.07$&  0.73$\pm 0.06$ &$1.00 \pm 0.07$& 0.52 $\pm$0.04& 
			0.15 $\pm$0.02 & $0.11\pm 0.01$ & $<0.1$ &$0.11 \pm 0.02$& 0.16 $^{+0.08}_{-0.06}$\\
Ne   \dotfill& 0.20$^{+0.07}_{-0.08}$&$0.85 \pm 0.08$ & 0.76$\pm$0.07 &$0.71 \pm0.07$& 0.18$\pm$0.03&
	   0.15$\pm 0.03$ &  $0.14 \pm 0.03 $& $<0.1$ &$0.12 \pm 0.02$& 0.21 $^{+0.1}_{-0.08}$ \\
Mg \dotfill & 0.12 $^{+0.06}_{-0.07}$&$0.51 \pm0.07$& 0.47$\pm 0.06$ &$0.34\pm 0.07$& 0.13 $\pm$0.03&
	   0.12$\pm 0.02$ & 0.16 $ \pm 0.03 $ & 0.12 $\pm$ 0.03 &$0.10 \pm 0.03$& 0.14 $\pm$ 0.02 \\
Si  \dotfill & $0.15 \pm 0.09$  &$0.70 \pm 0.1$& 0.8$\pm 0.1 $ &$1.5 \pm 0.2$& 1.1 $\pm$ 0.1&
	   1.7$\pm 0.1$ & 2.6$\pm 0.2$ & 1.3$\pm 0.1$ &$1.5 \pm 0.2$& 0.31$\pm 0.1$ \\
	   %
	%
Fe(=Ni)\dotfill& 0.14$^{+0.1}_{-0.04}$ &$0.53 \pm0.05$& 0.66 $\pm$0.04 &$ 0.82\pm 0.04$& 0.47$\pm$0.02& 
		0.66$\pm 0.02$ & 0.88 $\pm 0.02$ & 0.53$\pm 0.01$ &$0.56 \pm 0.02$& 0.19$\pm$0.1\\
$\tau [\times 10^{11}\mathrm{cm}^{-3} \mathrm{s}]$
	&0.99 $\pm$0.04 &$0.83 \pm 0.06$& 1.00  $\pm 0.04 $ &$1.3\pm 0.1$& 1.92$^{+0.1}_{-0.09}$
	&1.53 $^{+0.04}_{-0.12}$ & 1.33 $^{+0.07}_{-0.1}$ & 2.40 $^{+0.5}_{-0.3}$&$ 3.3^{+0.8}_{-0.9}$& 1.6 $\pm$0.05\\
 EM[$\times 10^{19}\mathrm{cm}^{-5}$] \dotfill &$4.20^{+0.3}_{-0.3}$ 
			&$0.05\pm0.01$&$0.046\pm0.009$&$0.03^{+0.01}_{-0.009}$&$0.10^{+0.003}_{-0.01}$
			&$0.077\pm0.003$&$0.076^{+0.005}_{-0.002}$&$0.178^{+0.01}_{-0.006}$&$0.11\pm0.01$
			& 0.37 $^{+0.3}_{-0.2}$\\
\hline
\multicolumn{11}{c} {Fixed abundance component} \\
\kTe [keV] \dotfill& $---$ &0.27$\pm$ 0.01 & 0.26 $\pm$0.01 &0.26$\pm 0.01$& 0.27$\pm0.01$
		  & 0.34 $\pm$0.01 & 0.28 $\pm$0.01 & 0.27$\pm$0.01 &0.33$\pm0.01$& $---$\\
		  
Abundances  \dotfill& $---$ & \multicolumn{8}{c}{(fixed to those determined for the NE rim of the Cygnus Loop)$^\dagger$} & $---$\\

$\tau [\times 10^{11}\mathrm{cm}^{-3} \mathrm{s}]$ \dotfill
	&$---$ &$0.56\pm 0.04$& 0.97 $\pm 0.04 $ &$1.02\pm0.07$& 0.72 $^{+0.04}_{-0.02}$
	&0.54 $^{+0.02}_{-0.03}$ & 0.92 $^{+0.08}_{-0.06}$ & 1.49$\pm$0.08&$0.71\pm0.04$&$---$ \\

EM[$\times 10^{19}\mathrm{cm}^{-5}$] \dotfill&$---$
	&$1.6\pm0.2$&$1.4\pm0.2$&$1.1\pm0.1$&$0.787^{+0.06}_{-0.1}$
	&$0.17^{+0.04}_{-0.007}$&$0.20^{+0.04}_{-0.01}$&$0.21^{+0.03}_{-0.01}$&$0.18^{+0.01}_{-0.02}$
        &$---$\\

\hline \hline
& & \\[-8pt]
\multicolumn{3}{@{}l@{}}{\hbox to 0pt{\parbox{140mm}{\footnotesize
\par\noindent
\footnotemark[$*$]Other elements are fixed to solar values.\\
The values of abundances are multiple of solar value.\\
The errors are in the range $\Delta\chi^2<$2.7 on one parameter\\
$^\dagger$C=0.27, N=0.11, O=0.11, Ne=0.21, Mg=0.17, Si=0.34, S=0.17, and Fe (=Ni)=0.20 (Uchida et al. 2006)
\par\noindent
}\hss}}
\label{param}
\end{tabular}
\end{center}
\end{sidewaystable}

\clearpage
\begin{thebibliography}{}
\bibitem[Anders \& Grevesse 1989]{Anders1989}
        Anders, E., \& Grevesse, N. 1989, Acta, 53, 197
\bibitem[Aschenbach \& Leahy 1999]{Aschenbach1999}
	Aschenbach, B. \& Leahy 1999, A\&A, 341, 602			 
\bibitem[Blair et al.\ 2005]{Blair2005}
        Blair, W. P., Sankrit, R., \& Raymond, J. C. 2005, AJ, 129, 2268
\bibitem[Borkowski et al.\ 2001]{Borkowski2001}
        Borkowski, K. J., Lyerly W. J., \& Reynolds, S. P. 2001,
        ApJ, 548, 820
\bibitem[Burrows et al.\ 2007]{Burrows2007}
	Burrows, A., Livne, E., Dessart, L., Ott, C. D., Murphy,
	 J. 2007, ApJ, 655, 416 
\bibitem[Decourchelle et al.\ 2001]{Decourchelle2001}
        Decourchelle, A. et al.\ 2001, A\&A, 365, L218
\bibitem[Dopita et al.\ 1977]{Dopita1977}
        Dopita, M. A., Mathewson, D. S., and Ford, V. L. 1977 ApJ, 214, 179
\bibitem[Hughes et al.\ 2000]{Hughes2000}
	Hughes, J. P., Rakowski, C. E., Burrows, D. N., Slane,
	 P. O. 2000, ApJ, 528, L109
\bibitem[Inoue et al.\ 1979]{Inoue1979}
        Inoue, H., Koyama, K., Matsuoka, M., Ohashi, T., Tanaka, Y.,
        Tsunemi, H. 1980 ApJ, 238, 886
\bibitem[Ishisaki et al.\ 2007]{Ishisaki2007}
        Ishisaki, Y., et al.\ 2007, PASJ, 59, S113
\bibitem[Katsuda \& Tsunemi 2007]{Katsuda2007}
	Katsuda, S. \& Tsunemi, H. 2007, Adv. Space Res, 41, 389
\bibitem[Katsuda et al. 2008]{katsuda2008}
	Katsuda, S., Tsunemi, H., Miyata, E., Mori, K., Namiki, M., Nemes N., Miller E. D. 2008, PASJ, 60S, 107
\bibitem[Koyama et al.\ 2007]{Koyama2007}
	Koyama, K. et al. 2007, PASJ, 59S, 23
\bibitem[Leonard et al.\ 2006]{Leonard2006}
	Leonard D.C. et al.\ 2006, Nature, 440, 505
\bibitem[Levenson et al.\ 1997]{Levenson1997}
        Levenson, N. A. et al. 1997, ApJ, 484, 304
\bibitem[Levenson et al.\ 1999]{Levenson1999}
        Levenson, N. A., Graham, J. R., and Snowden, S. L. 1999, ApJ, 526, 874
\bibitem[Leahy 2004]{Leahy2004}
	Leahy, D. A. 2004, MNRAS, 351, 385
\bibitem[Miyata et al.\ 1994]{Miyata1994}
        Miyata, E., Tsunemi, H., Pisarki, R., and Kissel, S. E. 1994,
          PASJ, 46, L101
\bibitem[Miyata et al.\ 1998]{Miyata1998}
        Miyata, E., Tsunemi, H., Kohmura, T., Suzuki, S., and Kumagai,
        S. 1998, PASJ, 50, 257
\bibitem[Miyata et al.\ 2000]{Miyata2000}
	Miyata, E., Tsunemi, H., Koyama, K., and Ishisaki, Y. 2000,
	 Adv. Space Res., 25, 555
\bibitem[Miyata et al.\ 2001]{Miyata2001}
	Miyata, E., Ohta, K., Torii, K., Takeshima, T., Tsunemi, H.,
	 Hasegawa, T., and Hashimoto, Y. 2001, ApJ, 550, 1023
\bibitem[Miyata et al.\ 2007]{Miyata2007}
        Miyata, E., Katsuda, S., Tsunemi, H., Hughes, J. P., Kokubun,
        M., and Porter, F. S. 2007, PASJ, 59S, 163
\bibitem[Morrison \& McCammon 1983]{Morrison1983}
        Morrison, R., \& McCammon, D. 1983, ApJ, 270, 119 
\bibitem[Parker et al.\ 2008]{park}
	Parker R. A. R. 1967, ApJ, 149, 363
\bibitem[Prigozhin et al.\ 2008]{sci}
	Prigozhin, G., Burke, B., Bautz, M., Kissel, S., LaMarr, B. 2008
	IEEE Transactions on Electron Devices, vol. 55
\bibitem[Rauscher et al.\ 2002]{Rauscher2002}
	Rauscher, T., Heger, A., Hoffman, R. D., Woosley, S. E. 2002,
	 ApJ, 576, 323
\bibitem[Thielemann et al.\ 1996]{Thielemann1996}
	Thilemann, F-K., Nomoto, K., \& Hashimoto, M. 1996, ApJ, 460, 408
\bibitem[Tsunemi et al.\ 1999]{Tsunemi1999}
        Tsunemi, H., Miyata, E., \& Aschenbach, B. 1999, PASJ, 51, 711        
\bibitem[Tsunemi et al.\ 2007]{Tsunemi2007}
	Tsunemi, H., Katsuda, S., Miyata, M., Nemes, N., Eric., M. D. 2007, ApJ, 671, 1717
	
\bibitem[Uchida et al.\ 2006]{Uchida2006}
	Uchida, H. Katsuda, S., Miyata, E., Tsunemi, H., Hughes, J. P.,
	 Kokubun, M., \& Porter, F. S. 2006, Suzaku Conference in Kyoto
\bibitem[Uchida et al.\ 2008]{Uchida2008}
	 Uchida, H. Tsunemi, H., Katsuda, S., Kimura, M. 2008 ApJ accepted
\bibitem[Woosley \& Weaver 1995]{Woosley1995}
	Woosley, S. E. \& Weaver, T. A. 1995, ApJS, 101, 181 

\end{thebibliography}
\end{document}